\documentclass[twocolumn, trackchanges]{aastex61} 
\usepackage{csvsimple}
\usepackage{multirow,bigdelim}
\usepackage{amsmath}

\newcommand{\msun}{\mbox{$M_\odot$}} 
\newcommand{\lsun}{\mbox{$L_\odot$}}
\newcommand{\rsun}{\mbox{$R_\odot$}}

\newcommand{\mearth}{M$_\oplus$}

\newcommand{\ms}{\mbox{m s$^{-1}$}}
\newcommand{\cms}{\mbox{cm s$^{-1}$}}
\newcommand{\kms}{\mbox{km s$^{-1}$}}

\newcommand{\teff}{${\rm T_{eff}}$}

\usepackage{colortbl}
\definecolor{tablegray}{rgb}{0.89, 0.89, 0.89}

\begin{document}

\title{EXPRES. II. Searching for Planets Around Active Stars: A Case Study of HD~101501}
\correspondingauthor{Samuel H. C. Cabot}

\email{sam.cabot@yale.edu}

\author[0000-0001-9749-6150]{Samuel H. C. Cabot}
\affil{Yale University, 52 Hillhouse Avenue, New Haven, CT 06511, USA}

\author[0000-0002-9288-3482]{Rachael M.\ Roettenbacher}
\affil{Yale Center for Astronomy and Astrophysics, Yale University, 46 Hillhouse Avenue, New Haven, CT 06511, USA}

\author[0000-0003-4155-8513]{Gregory W.\ Henry}
\affil{Tennessee State University, Center of Excellence in Information Systems, Nashville, TN 37203, USA}

\author[0000-0002-3852-3590]{Lily Zhao}
\affil{Yale University, 52 Hillhouse Avenue, New Haven, CT 06511, USA}

\author{Robert O.\ Harmon}
\affil{Ohio Wesleyan University, 61 S.\ Sandusky Street, Delaware, OH 43015, USA}

\author[0000-0003-2221-0861]{Debra A. Fischer}
\affil{Yale University, 52 Hillhouse Avenue, New Haven, CT 06511, USA}

\author[0000-0002-9873-1471]{John M. Brewer}
\affil{Department of Physics and Astronomy, San Francisco State University, 1600 Holloway Ave, San Francisco, CA 94132, USA}

\author[0000-0003-4450-0368]{Joe Llama}
\affil{Lowell Observatory, 1400 W. Mars Hill Rd., Flagstaff, AZ 86001, USA}

\author[0000-0003-2168-0191]{Ryan R. Petersburg}
\affil{Yale University, 52 Hillhouse Avenue, New Haven, CT 06511, USA}

\author[0000-0002-4974-687X]{Andrew E. Szymkowiak}
\affil{Yale University, 52 Hillhouse Avenue, New Haven, CT 06511, USA}

\begin{abstract}
By controlling instrumental errors to below 10 \cms, the EXtreme PREcision Spectrograph (EXPRES) allows for a more insightful study of photospheric velocities that can mask weak Keplerian signals. Gaussian Processes (GP) have become a standard tool for modeling correlated noise in radial velocity datasets. While GPs are constrained and motivated by physical properties of the star, in some cases they are still flexible enough to absorb unresolved Keplerian signals. We { apply GP regression to} EXPRES radial velocity measurements of the 3.5 Gyr old chromospherically active Sun-like star, HD~101501. We obtain tight constraints on the stellar rotation period and the evolution of spot distributions using { 28 seasons of ground-based photometry, as well as recent {\it TESS} data}. { Light curve inversion was carried out on both photometry datasets to reveal the spot distribution and spot evolution timescales on the star.} We find that the $> 5$ \ms\ rms radial velocity variations in HD~101501 are well-modeled with a GP stellar activity model without planets, yielding a residual rms scatter of 45 \cms. We carry out simulations, injecting and recovering signals with the GP framework, to demonstrate that high-cadence observations are required to { use GPs most efficiently to detect low-mass planets} around active stars like HD~101501. Sparse sampling prevents GPs from learning the correlated noise structure and can allow it to absorb prospective Keplerian signals. We quantify the moderate to high-cadence monitoring that provides the necessary information to disentangle photospheric features using GPs and to detect planets around active stars.

\end{abstract}


\section{Introduction} \label{sec:intro}

Radial velocity (RV) measurements have yielded numerous detections of exoplanetary systems via the gravitational interactions between planets and their host stars. In parallel, we have gained {a} deeper understanding of stellar physics and exoplanet populations. The improved technological capability of the latest generation of spectrographs enables the search for Earth analogues orbiting nearby stars in our galaxy. One such observing program focuses on Near-Infrared (NIR) observations of M-dwarfs with the CARMENES spectrograph \citep{Quirrenbach2010}; others observe low-mass stars in the optical using ultra-stable spectrographs such as ESPRESSO \citep{PePe2013, Mascareno2020} or EXPRES \citep{Jurgenson2016, Blackman2020, Petersburg2020}. Solar or sub-Solar mass stars are desirable RV targets because they exhibit greater reflex velocities from orbiting exoplanets. However, these host stars also have convective outer layers that contribute nuisance{d} signals in RV time series data. Along with high instrumental precision, it is of paramount importance to disentangle the Keplerian RV signal of an exoplanet from stellar activity signals. 

Significant effort has been dedicated toward theoretical or empirical modeling of the stellar activity RV contribution. One example is a simple spot model \citep{Aigrain2012}, which estimates the RV signature based on simultaneous photometric monitoring. Recent studies employ { quasiperiodic} Gaussian Processes (GPs) and moving-average methods to capture more complex correlated noise in RV time series \citep{Tuomi2013, Haywood2014}. These techniques force the model to be correlated on specific timescales, but uncorrelated after a parameterized decay timescale. Stellar activity may be probed, in part, by certain indicators and proxies derived from spectra. The Cross-Correlation Function (CCF) between a stellar spectrum and model template provides a couple of indicators \citep{Queloz2001, Queloz2009}. For example, the Bisector Inverse Slope  \citep[BIS;][]{Toner1988} probes granulation blueshift as a function of increasing height in the photosphere. Spots break the symmetrical, rotationally-broadened line profile as they move { across the stellar surface}, and produce variations in the CCF Full-Width at Half-Maximum (FWHM) \citep{Figueira2013}. Emission in cores of {c}alcium {\sc II} H\&K lines (denoted $\log R'_{\rm HK}$) probes chromospheric activity \citep{Saar1998, Cincunegui2007}. The H$\alpha$ core equivalent width is correlated with the overall photometric flux and used as a proxy in the simple spot model \citep{Giguere2016}. Stellar activity may also be isolated by its impact on individual lines \citep{Davis2017, Dumusque2018, Cretignier2020}. However, despite these diagnostic advances, there is no robust methodology for consistently distinguishing low-mass planetary signals from stellar activity in RV datasets \citep{Dumusque2017}. 

Higher fidelity data acquired by new state-of-the-art spectrographs can reach $\lesssim 30 \: \cms$ measurement precision \citep{Mascareno2020, Brewer2020} and might hold clues to solving this longstanding problem. We thus turn our attention to HD~101501 (61 UMa), which provides an exemplary case of stellar activity that dominates an RV time series. The star is bright ($V=5.31$) and Sun-like (G8V) \citep{Boyajian2012} and is a target in the EXPRES observing program \citep{Brewer2020}. Historically, the star has been used as a standard for spectral classification of stars \citep{Johnson1953} and is commonly included in population studies of Sun-like stars \citep{Duquennoy1991, Valenti2005, Fischer2005}. HD~101501 has a rotation period of $\sim 16$ days \citep{Donahue1996} and no confirmed companions. { \citet{Fischer2014} published RVs for HD 101501 from the {\it Lick} Observatory Hamilton spectrograph with an rms of 13.48 \ms\ and \citet{Howard2016} show a similar rms of 13.12 \ms\ from combined {\it Lick} and {\it Keck} data.} This RV scatter is large enough to mask Keplerian signals of planets with mass $\lesssim 100$ \mearth. Active, young stars are usually off-limits for RV observing programs, which explains the lack of HD~101501 observations in HARPS or HARPS-N archival databases. However, these properties make the star a fascinating case study for in-depth characterization, as well as a testbed for methods which reduce correlated RV noise. 

As follows we present new high-precision RVs of HD~101501 with simultaneous photometry. Details regarding the data are in Section~\ref{sec:obs}. We employ Gaussian Processes for modelling stellar activity. In Section~\ref{sec:gp} we review the GP method { and} its application to our dataset, { as well as} present benchmarks on archival data. Section~\ref{sec:res} contains the results of our planet search of HD~101501, in which we compare multiple GP-based models and find that an activity-only (zero planet) model has the highest evidence. One of the limitations of the RV data is seasonal low-cadence over certain stretches of time, which hinders the GP from learning activity signals and distinguishing them from possible short-period Keplerians. We quantify the impact of cadence in Section~\ref{sec:cad} and recommend observing strategies for detecting planets around active stars. We find high-cadence is necessary for using GPs to detect low-amplitude planets around HD~101501 and other active stars. We perform a detailed characterization of stellar rotation and activity in Section~\ref{section:disc} by performing light curve inversion. Section~\ref{section:con} summarizes our results. 

\section{Observations} \label{sec:obs}

Our data consist of high-precision radial velocity measurements and simultaneous ground-based photometry. This combination is advantageous because it allows joint constraints on stellar activity between the two time series \citep{Haywood2014}. Details surrounding data acquisition and processing are as follows. We also analyze photometry from the Transiting Exoplanet Survey Satellite (\emph{TESS}), which partially overlaps the recent EXPRES RV data.

\subsection{Photometry with APT} \label{subsec:obsapt}

\begin{figure*}
    \centering
    \includegraphics[width=\linewidth]{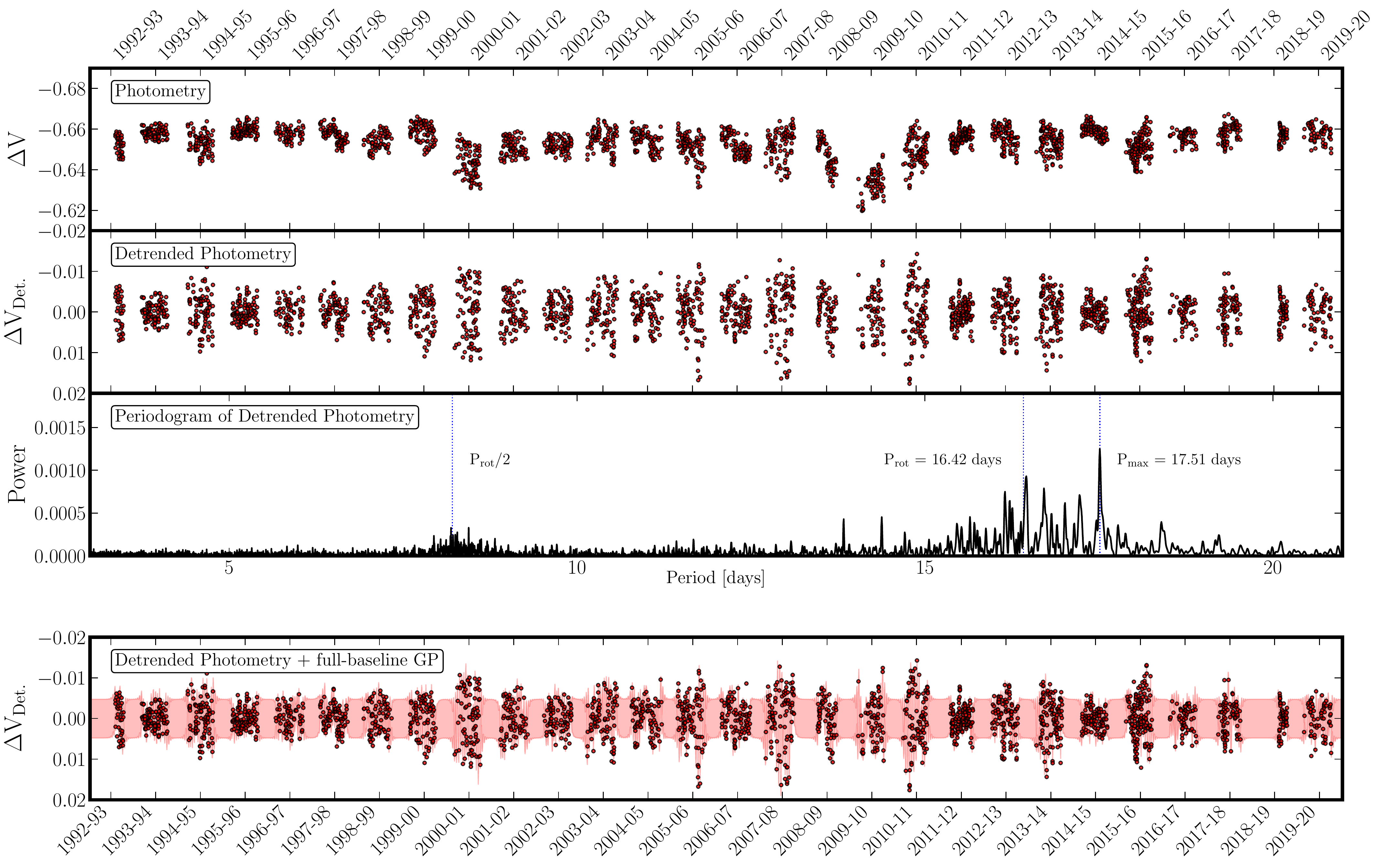}
    \caption{Full APT photometry dataset providing V-band monitoring of HD~101501. {\it Top Panel}: dataset in units of relative magnitude. {\it Second Panel}: dataset following detrending procedure described in the text. Long-term variations are removed. The scatter within each season reflects stellar activity modulated by the rotation period. {\it Third Panel}: Generalized Lomb-Scargale Periodogram of the detrended photometry. We mark the period of maximum power $P_{\rm max}$, our determined rotational period $P_{\rm rot}$ (see Section~\ref{section:disc}), and $P_{\rm rot}/2$. {\it Bottom Panel}: detrended photometry plotted with the full-baseline, \texttt{celerite}, quasiperiodic GP fit. The red band is the $1\sigma$ confidence interval of the GP. We infer the rotation period of the star as the {\it maximum a posteriori} value of the GP periodic timescale parameter, $P_{\rm rot} = P_{\rm GP}$ (details in Section~\ref{section:disc}).
    }
    \label{fig:aptfull}
    \vspace{5mm}
\end{figure*}

Fairborn Observatory's Automatic Photoelectric Telescopes  \citep[APTs{;}][]{Henry1999} observed HD~101501 for 28 seasons (spanning 18 April 1993 to 21 June 2020). Observations have about 1 day typical cadence for 6-7 months each year, totalling 2673 data points. The data display significant correlated structure arising from stellar activity in the target and have a standard deviation of 7.1 mmag in the V-band.  We focus on activity from spots, which are modulated by the rotational period, and remove long-term brightness variations from the light curve.  The long-term trend of the light curve that cannot be simply attributed to rotational modulation is removed by smoothing the light curve over 100 days and subtracting the resultant trend. The light curve standard deviation is reduced to 4.5 mmag, following detrending (Figure~\ref{fig:aptfull}). The periodogram has complex structure around 17 days, close to the 16.18 day rotation period estimated by \citet{Donahue1996}. The maximum power is at 17.51 days. 

\subsection{Photometry with TESS} \label{subsec:obstess}

\begin{figure*}
    \centering
    \includegraphics[width=\linewidth, trim={0.75cm 1.8cm 0.75cm 0.5cm},clip]{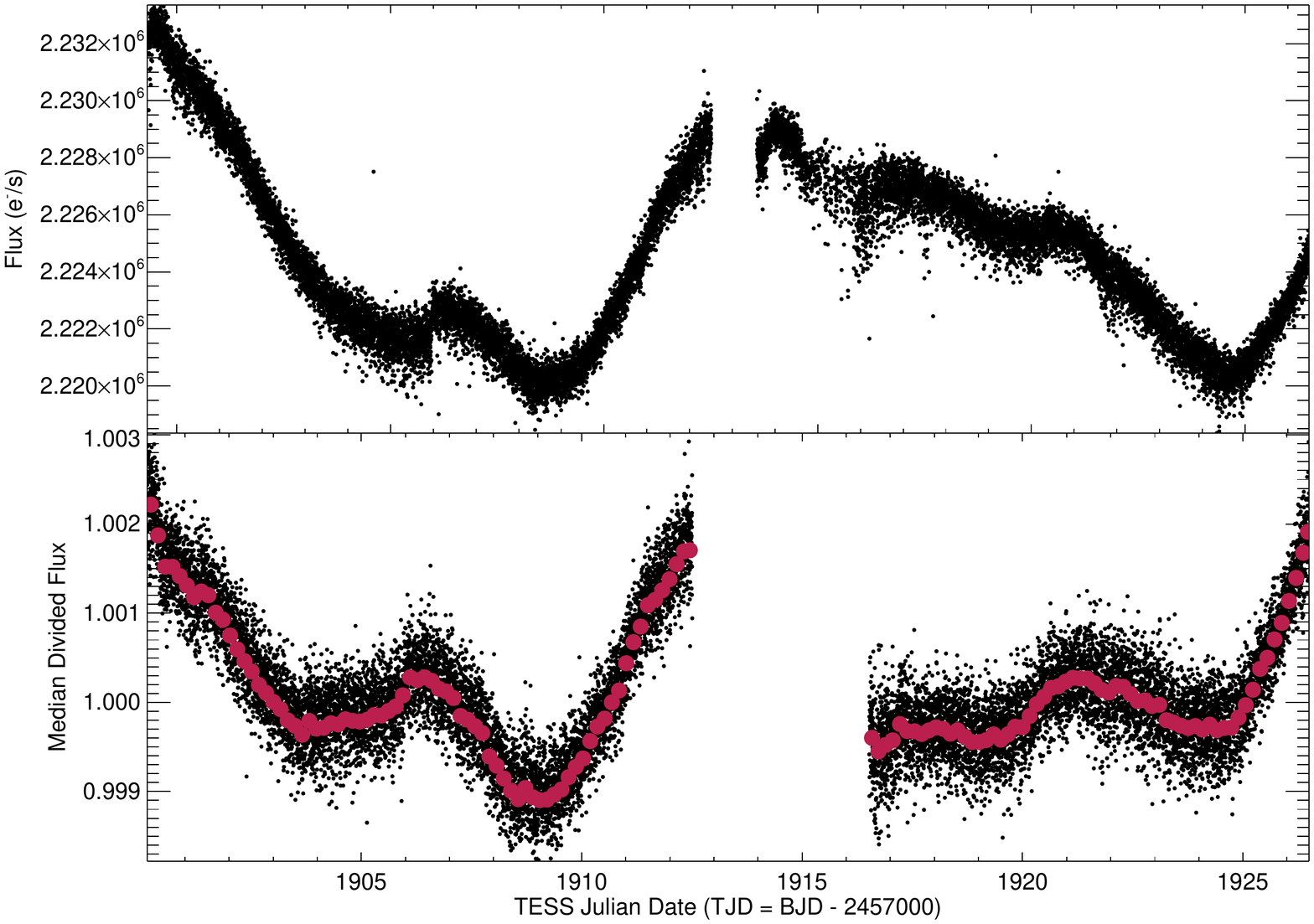}
    \caption{Sector 22 \emph{TESS} light curve.  Top:  The 2-minute SAP light curve provided by standard \emph{TESS} pipeline.  Bottom:  The \emph{TESS} light curve with the first five cotrending basis vectors removed.  Plotted in red are the average flux values for the binned light curve used in our inversions, see Section \ref{subsec:LI}.  When used in the inversions, these fluxes are normalized for each rotation, here they are overlaid for comparison. } 
    \label{fig:TESSLC}
\end{figure*}

The Transiting Exoplanet Survey Satellite \citep[TESS;][]{Ricker2014} observed HD~101501 during Sector 22 (18 February 2020 to 18 March 2020; Figure~\ref{fig:TESSLC}).  The 2-minute cadence, simple aperture photometry (SAP) light curve was used and the first 5 co-trending basis vectors were applied to remove instrumental signatures in order to preserve stellar astrophysics \citep[following the process used in][]{Roettenbacher2018}.  Both the light curve and co-trending basis vectors were obtained through the Barbara A.\ Mikulski Archive for Space Telescopes (MAST).

\subsection{Spectroscopy with EXPRES} \label{subsec:obsexpres}

We analyze 76 RVs (over 33 distinct nights) of HD~101501 obtained between 2018-2020. Data were collected with the Extreme PREcision Spectrometer (EXPRES) commissioned at the 4.3-m {\it Lowell Discovery Telescope} ({\it LDT}) \citep{Levine2012} (observing program: {\it The 100 Earths Survey}). EXPRES achieves $\sim 30$ \cms\ measurement precision for a pixel signal-to-noise ratio of $250$ at 5500~\AA. EXPRES has typical resolving power $R\sim137,500$ and spans { a wavelength range} of $380-780$ nm. More details regarding EXPRES may be found in recent studies investigating performance benchmarks and detailing the radial velocity extractions \citep{Petersburg2020, Blackman2020, Brewer2020}. We also extract activity indicators (CCF FWHM, CCF BIS and H$\alpha$ Equivalent Width) for each exposure, derived from the spectrum and CCF. {\it The 100 Earths Survey} targets chromospherically quiet stars without close-in gas giants in order to search for terrestrial planets. HD~101501 represents one of several additional, active stars observed for purposes of investigating and mitigating activity signals. Physical attributes of HD~101501 are listed in Table~\ref{tab:star}, { and summary statistics for the EXPRES exposures are in Table~\ref{tab:obs}.} { For convenience, we subtract the mean of the RV data (about $-5.55$ \kms) corresponding to the systemic velocity.}

\begin{table}
\setlength{\tabcolsep}{4pt}
\begin{tabular}{ l l l r }
\hline
  Property & Symbol & Units & Value$^*$ \\
\hline
  Visual Magnitude         & $V$           & mag.                  & $ 5.31 $ \\
  Distance                 & $ d $         & pc                    & $ 9.61 $ \\
  Effective Temperature    & $T_{\rm eff}$ & K                     & $ 5502 $ \\
  Surface Gravity          & $\log g$      & cm s$^{-2}$           & $ 4.52 $ \\
  Metallicity              & $[$Fe/H$]$    & $[$Fe/H$]_{\odot}$    & $ -0.04 $ \\
  Age                      & $t_{\rm age}$ & Gyr.                  & $ 3.5_{-2.2}^{+2.8} $ \\
  Proj. Rotation Speed & $v\sin i$     & \kms\                 & $ 2.2 $ \\
  Luminosity               & $\log$ $L_*$  & $\log$ \lsun          & $ -0.21 \pm 0.02 $ \\
  Radius                   & $R_*$         & \rsun                 & $ 0.86 \pm 0.02 $ \\
  Mass                     & $M_*$         & \msun                 & $ 0.90 \pm 0.12 $ \\
\hline
\end{tabular}
  \caption{Physical properties of HD~101501. $^*$All values and uncertainties are as reported by \citet{Brewer2016}.}
  \label{tab:star}
\end{table}

\begin{table}
\setlength{\tabcolsep}{2pt}
\begin{tabular}{ l l l r }
\hline
  Property & Symbol & Units & Value \\
\hline
  Number of Exposures       & N$_{\rm exp}$       & -         & 76 \\
  Number of Nights          & N$_{\rm night}$     & -         & 33 \\
  Time Baseline             & -                   & days      & 796 \\
  Average RV                & $\widehat{\rm RV}$  & \ms       & -5554.43 \\
  RV Root-Mean-Square       & rms                 & \ms       & 6.1 \\
  Median Measurement Uncertainty & $\tilde{\sigma}_n$    & \ms       & 0.38 \\
\hline
\end{tabular}
  \caption{Summary of EXPRES radial velocity measurements used in this study.}
  \label{tab:obs}
\end{table}

\section{Gaussian Processes for Modeling Stellar Activity} \label{sec:gp}

Before describing { the} GP framework, we review physical processes { within the star that are} responsible for RV variations. They may be categorized as follows \citep{Dumusque2012, Fischer2016, Cegla2019}: (1) p-mode acoustic oscillations in the convective envelope \citep{Chaplin2013}, which induce RV variations of order $\lesssim 1$ \ms\ on timescales of several minutes. Observing strategies can often average out and reduce this effect to within the instrument precision \citep{Dumusque2011}; (2) granulation cells consisting of rising hot gas and descending cool gas. Depending on the temperature of the gas, granulation induces a net blueshift \citep{Dravins1982} and variability of tens of \cms\ \citep{Schrijver2000} over the course of several minutes; (3) magnetic activity cycles like the solar cycle, which inhibit the convective blueshift \citep{Meunier2013}, typically on timescales of years; and (4) spots and faculae, which can create especially pernicious RV variations with amplitudes and periods commensurate with planetary signals \citep{Boisse2011}. The magnetic fields associated with spots and faculae inhibit convection in local regions on the star. The effect is modulated by the stellar rotation, and therefore can be confused for an exoplanet with an orbital period at a harmonic of the stellar rotation period \citep{Queloz2001}, especially in under-sampled radial velocity datasets. Starspot lifetimes vary from days to years \citep{Hall1994, Berdyugina2005} and hence induce a quasiperiodic RV variation. In some cases, the sinusoidal variation remains coherent over several months \citep{Robertson2020}. 

\subsection{GP Formalism}

\begin{figure}
    \centering
    \includegraphics[width=\linewidth]{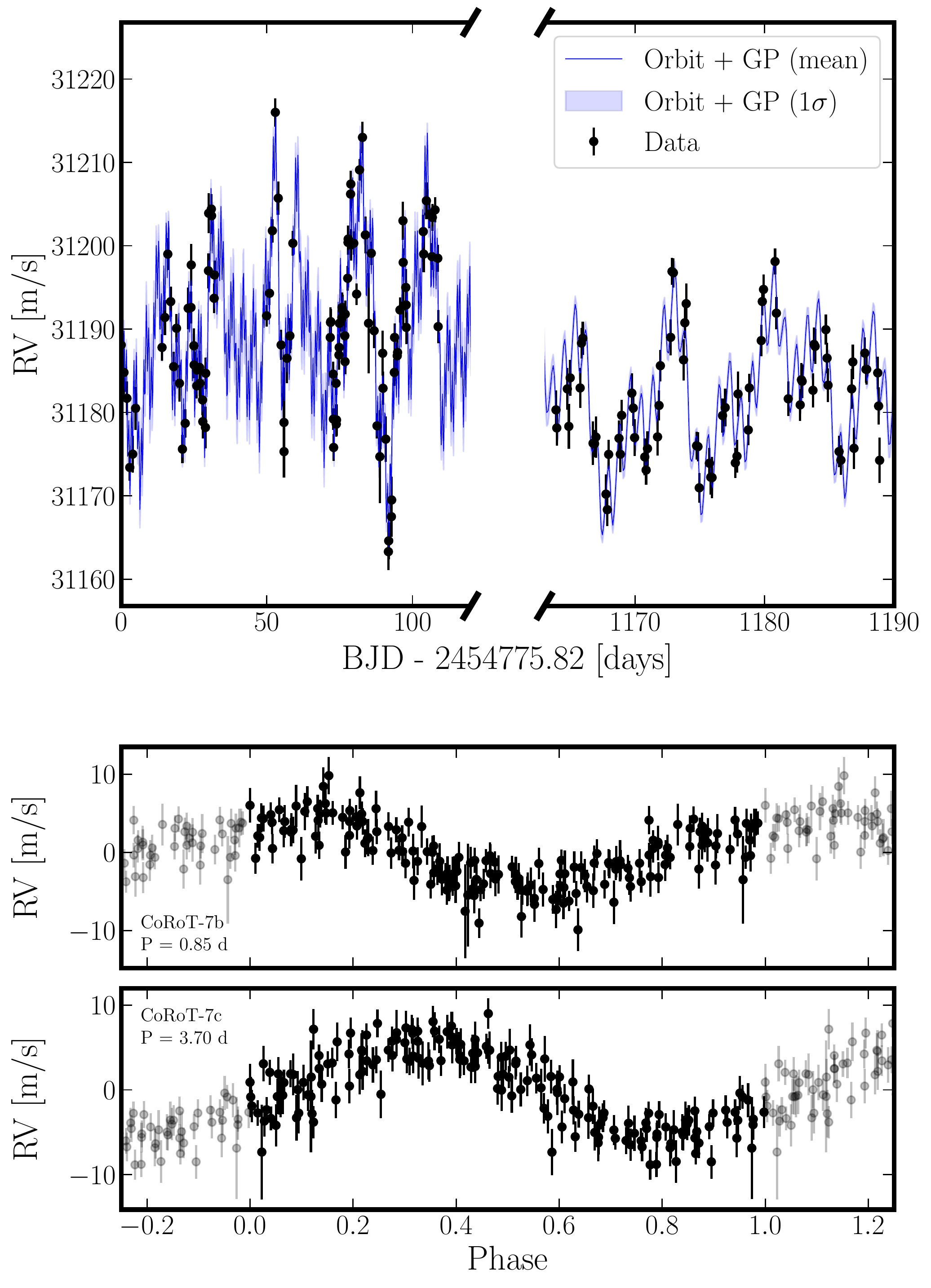}
    \caption{The orbits of CoRoT-7b and CoRoT-7c, retrieved by applying the \texttt{george} quasiperiodic GP + 2-Planet model to RV observations of the system. These detections are made on the same dataset analyzed by \citet{Faria2016}, and are consistent with their results, as well as results presented by \citet{Haywood2014}. The top panel shows the data and best-fit model, including the GP $1\sigma$ confidence interval. The bottom panels show, for each planet, the phase-folded orbit after subtracting the other planet's orbit and GP mean. 
    }
    \label{fig:f16}
\end{figure}

The stochastic nature of stellar activity makes it difficult to model analytically, and has motivated the use of GPs \citep{Haywood2014, Rajpaul2015}. A GP is a flexible model which assumes data points are drawn from a multivariate normal distribution \citep{Rasmussen2006}. Under a GP model, RV measurements ${y}$ at times ${t}$ have joint distribution
\begin{equation}
    p({y;t}) \sim \mathcal{N}(m({t}), K({t, t}) + \sigma_n^2I),
\end{equation}
where $m$ is a mean function and $K$ is a covariance matrix. The mean and covariance functions have hyperparameter vectors $\theta$ and $\phi$ respectively. The white noise term involves measurement uncertainties $\sigma_n$. GPs can serve as predictive models for estimating values and uncertainties at times between measurements. The agreement between a GP model and observed data may be quantified by the logarithm of the marginal likelihood
\begin{multline} \label{eqn:llh}
    \log \mathcal{L(\theta, \phi)} = -\frac{1}{2}{r}^\top(K + \sigma_n^2I)^{-1}{r} \\
    - \frac{1}{2}\log|K + \sigma_n^2I| - \frac{n}{2}\log2\pi,
\end{multline}
where the vector of residuals is
\begin{equation}
    {r} = {y} - m({t}).
\end{equation}
GPs have been used extensively in the literature for modeling correlated noise in RV datasets, and enabling detections of low-mass planets \citep[e.g.][]{Haywood2014, Rajpaul2015, Grunblatt2015, Cloutier2017, Faria2020, Mascareno2020, Benatti2020}. In these cases, $m$ as defined above takes the form of a Keplerian signal or sum of multiple Keplerians. The Keplerians increase $\log \mathcal{L(\theta, \phi)}$ by decreasing the residuals in ${r}$ and changing the optimal hyperparameters. Some restrictions in the GP framework make it more appropriate for modelling stellar noise in RV datasets, and prevent it from absorbing planetary signals; namely, a quasiperiodic covariance function
\begin{equation} 
\label{eqn:qpc}
    K_{ij} = a^2\exp\Big[-\frac{|t_i - t_j|^2}{\lambda_e^2} - \frac{\sin^2(\pi|t_i-t_j|/P_{\rm GP})}{\lambda_p^2}\Big]
\end{equation}
which is the product of squared exponential and sinusoidal covariance functions. The hyperparameters $\phi = \{a^2, \lambda_e, \lambda_p, P_{\rm GP}\}$ correspond to the magnitude of covariance, a decay parameter for the overall GP evolution, a dimensionless { smoothing parameter} for the periodic component, and the period of oscillations, respectively. This choice is motivated by the underlying physics when $P_{\rm GP}$ equals $P_{\rm rot}$, the rotation period of the star, and $\lambda_e$ is related to the typical lifetimes of spots; however direct interpretations of hyperparameters beyond the rotation period are tenuous \citep{Rajpaul2015}. Often a jitter parameter $s$ is added in quadrature with measurement uncertainties \citep{Grunblatt2015}.
\citet{Foreman-Mackey2017} show the covariance function,
\begin{equation}
    K_{ij} = \frac{B}{2+C}e^{-|t_i-t_j|/L}\Big[\cos{\frac{2\pi|t_i-t_j|}{P_{\rm GP}}} + (1 + C)\Big],
    \label{eqn:k2}
\end{equation}
behaves similar to Equation~\ref{eqn:qpc}, and allows faster matrix inversion (Equation~\ref{eqn:llh}). It has been used in recent RV studies \citep{Mascareno2020, Robertson2020} and compared to other available kernels \citep{Espinoza2020}. Hyperparameters $\phi = \{B, C, L, P_{\rm GP}\}$ correspond to the magnitude of covariance, weighting of the sinusoidal term, decay parameter, and period, respectively. Most studies which use GPs to model correlated RV noise adopt one of the above two covariance functions, and for our analyses we use the \texttt{george} implementation \citep{Ambikasaran2015} for Equation~\ref{eqn:qpc} and the \texttt{celerite} implementation \citep{Foreman-Mackey2017} for Equation~\ref{eqn:k2}. We briefly note a few differences between the kernels. The \texttt{celerite} GP is not mean-square differentiable \citep{Rasmussen2006} and less smooth than the \texttt{george} GP. Also the covariance decreases faster on short timescales compared to the \texttt{george} GP for equal $\lambda_e = L$. RV variations { due to the star} are stochastic on many different timescales and are not necessarily a smooth or coherent process. However, the actual power-spectrum of { high-frequency variations} will depend on the RV signatures of granulation and oscillations, which are not well understood. Given its smoothness, Equation~\ref{eqn:qpc} is a more attractive model for activity associated with faculae and spots, which themselves evolve on timescales comparable to the rotation period. We applied { the} GP framework with both kernels on the identical CoRoT-7 dataset analyzed by \citet{Faria2016}. The GP + 2-Planet model favored use of the \texttt{george} GP over the \texttt{celerite} GP with $\Delta\ln{\mathcal Z} \approx 8$ (this metric is described in the following subsection). In both cases we found that the sampler tended to converge to an alias of the 0.85 day period planet at $\sim 6$ days, but that the shorter period planet is visible in the residuals and periodogram following subtraction of the $3.7$ day planet and GP. The correct orbital period was retrieved after imposing a prior restricted to orbital periods $P<5$ days. Figure~\ref{fig:f16} shows our best-fit. Moving forward we adopt the \texttt{george} GP for modelling RV noise.

\subsection{Application to EXPRES RVs and APT Light Curve}

{ We couple the} GP framework { with} the importance nested sampling algorithm \texttt{MultiNest} \citep{Feroz2008, Feroz2009, Feroz2019} via the \texttt{PyMultiNest} implementation \citep{Buchner2014}. The parameter space includes GP hyperparameters $\{a^2, \lambda_e, \lambda_p, P_{\rm GP}\}$, jitter parameter $s$, systemic { offset} $\gamma$, and orbital parameters $\{K_s, \phi_0, P, \omega, e\}$ for $N$ planets, corresponding to semi-amplitude, phase of first epoch, orbital period, longitude of periastron, and eccentricity, respectively. We adopt $\phi_0$ as a boundary condition instead of time of periastron ($T_p$) since there are no degeneracies in the $[0, 2\pi]$ prior. \texttt{MultiNest} returns the log evidence of the model, $\ln{\mathcal Z}$, which may be used for model comparison for $N = \{0, 1, 2, 3, ...\}$ planets. Given dataset $\mathbf{d}$ and a model ${\mathcal M}$ parametrized by a vector $\theta$, the Bayesian evidence is defined as
\begin{equation}
    {\mathcal Z} \equiv p(\mathbf{d}|{\mathcal M}) = \int p(\mathbf{d}|{\theta}, {\mathcal M})  p({\theta} | {\mathcal M}) d{\theta}
\end{equation}
\citep[Bayesian evidence in the context of RV analyses is discussed at length by][]{Nelson2020}.
Assuming equal prior odds on given models ${\mathcal M_0}$ and ${\mathcal M_1}$ (e.g. zero planets and one planet, respectively), it is generally agreed that a difference in corresponding evidences $\ln{\mathcal Z_1} - \ln{\mathcal Z_0} \geq 5$ indicates ${\mathcal M_1}$ is strongly preferred over ${\mathcal M_0}$ \citep{Kass1995}. { Our model evidences and uncertainties correspond to the median and standard deviation of evidences from five runs of the sampler. Repeated runs are known to provide more reliable uncertainty estimates than single-run output \citep{Nelson2020}.} Our choices of priors are listed in Table~\ref{tab:priors}. { The priors on orbital parameters represent our expectations for what could be detected in the data, given the sparse sampling and large fluctuations}.

\begin{table}
\renewcommand{\arraystretch}{0.825}
\setlength{\tabcolsep}{2pt}
\begin{tabular}{ l l l r }
\hline
  Parameter & Definition & Units & Distribution \\
\hline
\hline
GP & & & \\
\hline
  $a$                & Amplitude                & \ms  & ${\mathcal LU}(0.1, 1000)$   \\
  $\lambda_e$        & Decay Timescale          & days & $\delta({18.74})$ \\
  $P_{\rm GP}$       & Periodic Timescale       & days & $\delta({16.28})$ \\
  $\lambda_p$        & { Smoothing Parameter}& -    & $\delta(0.76)$ \\
\hline
Global & & & \\
\hline
  $s$                & Jitter                   & \ms  & ${\mathcal U}(0.01, 5)$ \\
  $\gamma$           & Systemic { Offset}    & \ms & ${\mathcal U}({-20, 20})$ \\
\hline
Orbital & & & \\
\hline
  $K_s$              & Semi-amplitude           & \ms  & ${\mathcal LU}({0.1}, 10)$ \\
  $\phi_0$           & Phase of First Epoch     & rad. & ${\mathcal U}(0, 2\pi)$ \\
  $P$                & Period                   & days & ${\mathcal LU}(0.5, 20)$ \\
  $\omega$           & Longitude of Pericenter  & rad. & ${\mathcal U}(0, 2\pi)$ \\
  $e$                & Eccentricity             & -    & ${\mathcal LU}(0, 0.99)$ \\
\hline
\end{tabular}
  \caption{Priors on parameters for the light curve-conditioned, 1-Planet RV model. We sample \texttt{george} GP covariance parameters (Equation~\ref{eqn:qpc}), global parameters including a jitter term and RV offset, and orbital parameters for a single planet. The table lists each parameter, its corresponding units, and prior distribution. For uniform (${\mathcal U}$) and log-uniform (${\mathcal LU}$) distributions, we specify upper and lower bounds. For this model, three of the four GP parameters are fixed (denoted by $\delta$) to values inferred from photometry.}
  \label{tab:priors}
\end{table}

We perform a similar analysis as \citet{Haywood2014} by conditioning GP hyperparameters based on simultaneous photometry. We fit a \texttt{george} GP model to the most recent three seasons of photometry (2018-2020), including a jitter parameter and constant offset. Note, the matrix inversion becomes intractable for many data points, so we restrict this step to the timeframe overlapping with EXPRES RV observations, neglecting pre-2018 data. { We use log-uniform priors spanning multiple orders of magnitude on all parameters with two exceptions: the constant offset is drawn from a uniform prior, and the periodic timescale is bounded by 12 days and 22 days. This latter constraint was chosen to prevent convergence on a harmonic or multiple of the rotation period.} Afterwards, when we use a GP model to fit the RVs, the hyperparameters $\lambda_e, \lambda_p$, and $P_{\rm GP}$ are fixed to the {\it maximum a posteriori} (MAP) values from photometry fitting. The amplitude $a$ is allowed to be different. Fixing model parameters decreases the sampling dimensionality and computation time. It also informs the model in case the RVs alone are insufficient to constrain the stellar rotation period. It is expected that GP fits to RV and photometry time series should have similar $\lambda_e$, $\lambda_p$ and $P_{\rm GP}$, as demonstrated by \citet{Kosiarek2020} with solar data (see their Figure 9).

\section{Results} \label{sec:res}

We now present results of GP model fits to the photometry and RVs, which includes searching for planetary companions around HD~101501. 

\subsection{Photometric Contraints} \label{subsec:photo}

\begin{figure*}
    \centering
    \includegraphics[width=\linewidth]{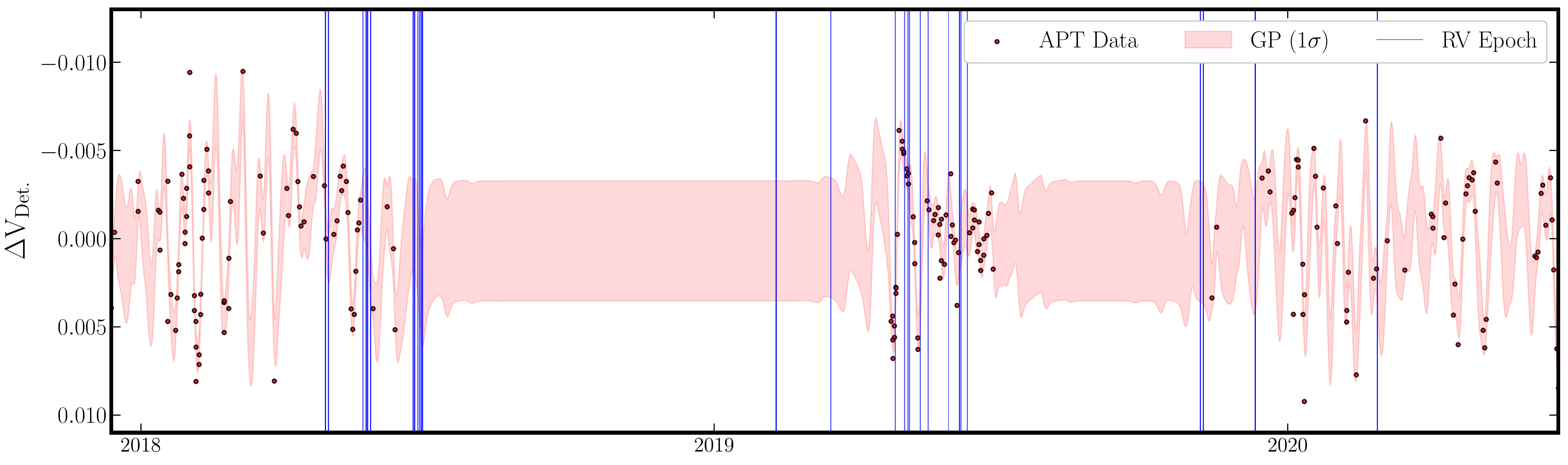}
    \caption{Detrended APT photometry of HD~101501 for 2018, 2019, and 2020 observing seasons. The red shaded region depicts the $1\sigma$ confidence region around the regressed \texttt{george} GP mean. { The blue lines} denote epochs of RV measurements.}
    \label{fig:aptrecent}
    \vspace{5mm}
\end{figure*}

\begin{figure}
    \centering
    \includegraphics[width=\linewidth]{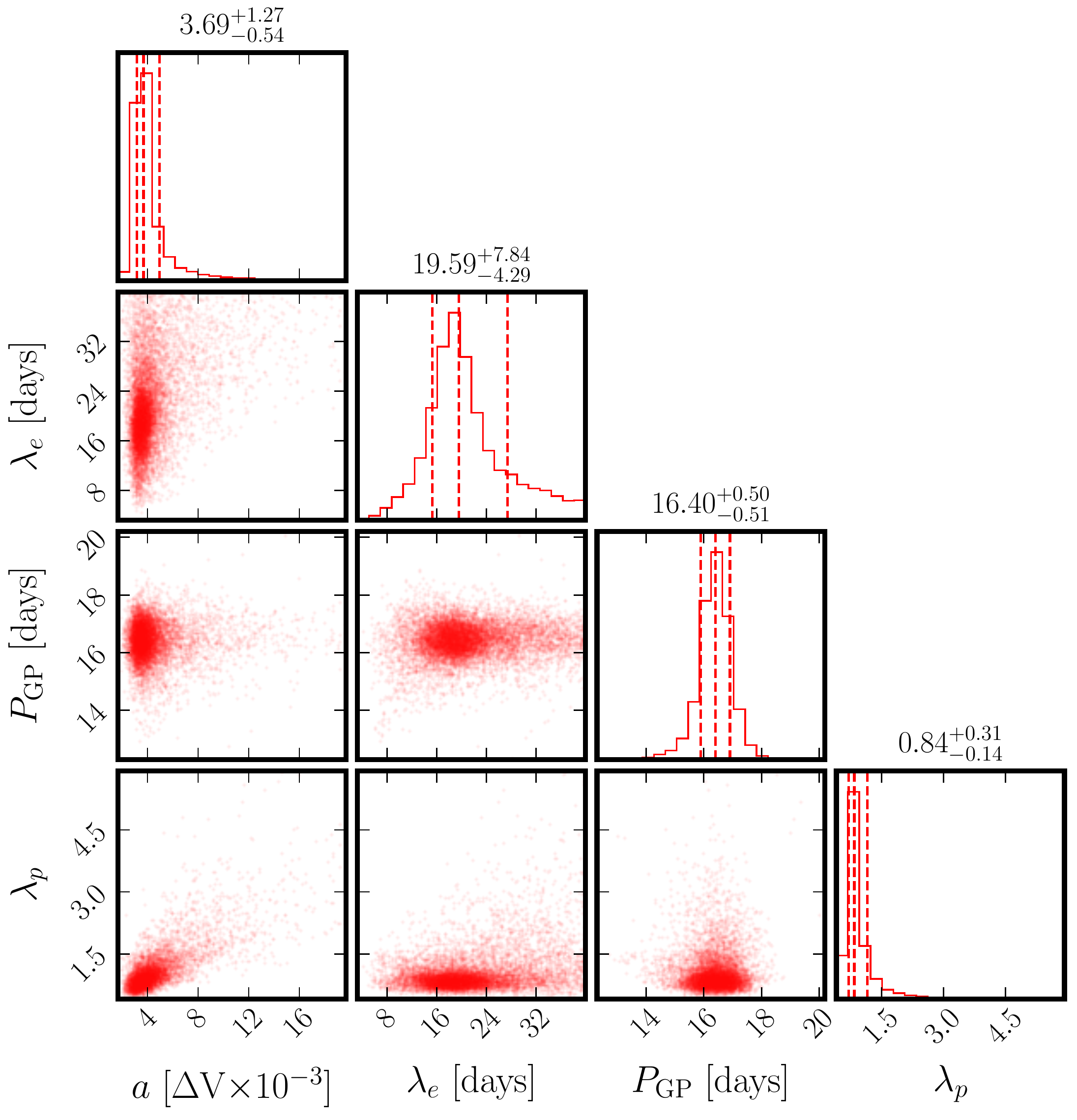}
    \caption{Marginalized posterior distributions of hyperparameters for the \texttt{george} GP model regressed to the 2018-2020 APT photometry. Samples were obtained via importance nested sampling of the posterior distribution, as described in the text. Median values are plotted above the 1-D histograms, where uncertainties correspond to $16^{\rm th}$ and $84^{\rm th}$ percentiles. These quantiles are marked by red dashed lines.}
    \label{fig:photocorner}
\end{figure}

In fitting a \texttt{george} GP model to 2018-2020 photometry, we obtain MAP hyperparameter values $a={3.40}$ mmag., $\lambda_e={18.74}$ days, $P_{\rm GP}={16.28}$ days, and $\lambda_p={0.76}$. { The \texttt{george} GP with MAP hyperparameters is plotted against the recent photometry data in Figure~\ref{fig:aptrecent}.} The joint-distributions between GP hyperparameters and their marginalized histograms are shown in Figure~\ref{fig:photocorner}. All of the hyperparameters are constrained around well-defined peaks. { Both $\lambda_{e}$ and $\lambda_{p}$ have an effect on whether the GP rapidly varies or gradually changes \citep{Rasmussen2006}. Larger $\lambda_{e}$ allows the GP to repeat itself more times before it loses coherence. Larger $\lambda_{p}$ forces the repeating signal to be smoother, whereas smaller $\lambda_{p}$ allows more fine-structure.} { While not strictly enforced by our priors, the evolutionary timescale converged to a value larger than the periodic timescale ($\lambda_{e} > P_{\rm GP}$). This is a realistic constraint in regressing quasiperiodic GPs to photometry \citep{Kosiarek2020}.}

\begin{figure*}
    \centering
    \includegraphics[width=\linewidth]{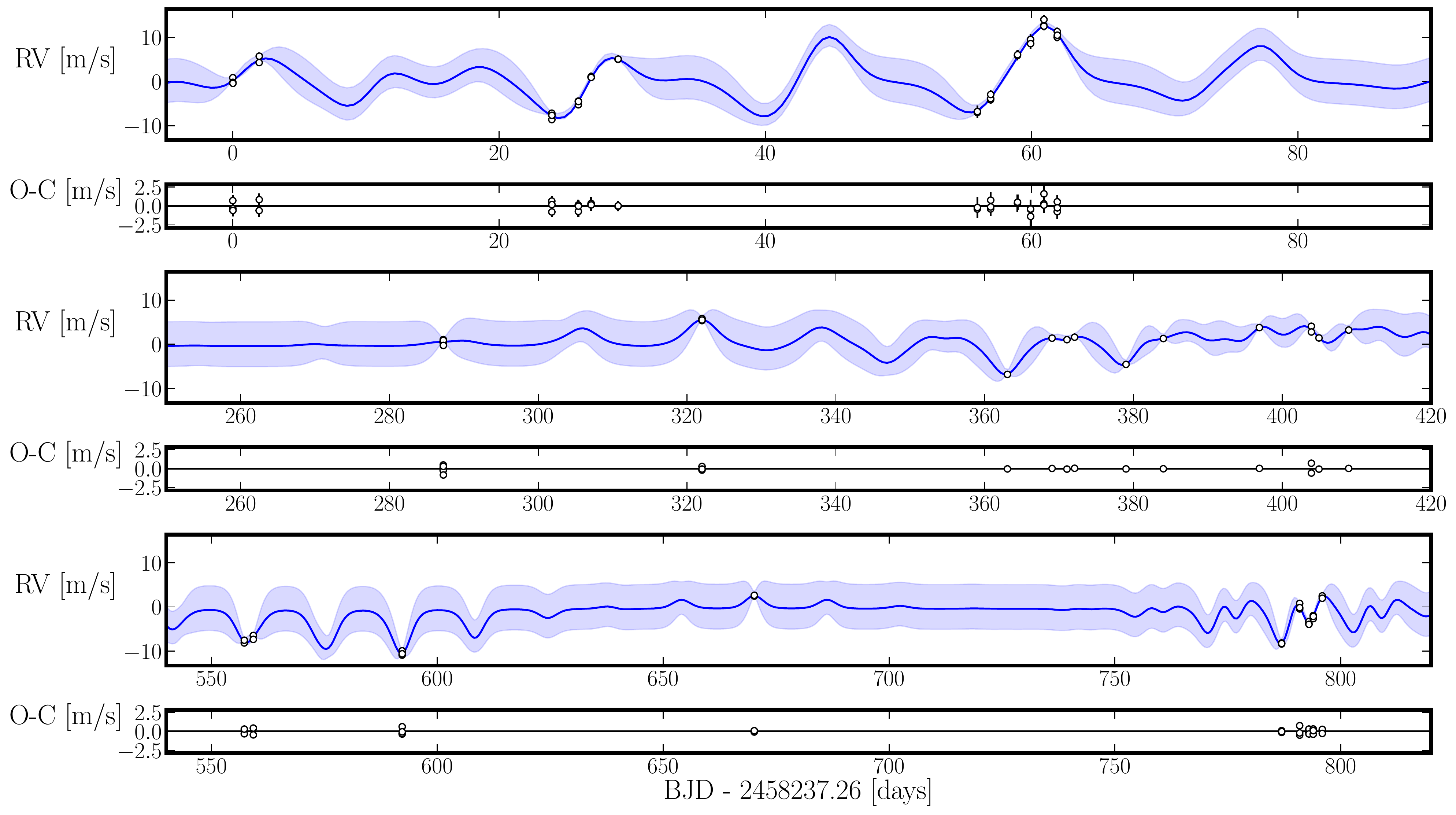}
    \caption{GP Only fit to EXPRES RV measurements of HD~101501. The \texttt{george} quasiperiodic GP has three fixed hyperparameters based on the photometry data. The time series is divided into three sections for clarity. The GP mean (plus systemic offset) is denoted by the blue line, and the $1\sigma$ confidence interval on the GP is the light-blue shaded region. The residuals are plotted beneath each panel, with rms scatter 0.45 $\ms$.}
    \label{fig:1p}
\end{figure*}

\subsection{RV Characterization and Planet Search} \label{subsec:fitrv}

The 0-Planet (GP-only) model is most favored at a log evidence of ${\ln {\mathcal Z} = -149.26 \pm 0.15}$, compared to the GP + 1-Planet model at ${\ln {\mathcal Z} = -149.72 \pm 0.06}$. We additionally restricted the Keplerian to a sinusoid (circular orbit) by fixing $\omega=0$ and $e=0$. This model returned ${\ln {\mathcal Z} = -149.75 \pm 0.18}$. { For the 1-Planet model, the MAP planet period is at 15.24 days, which is close to the adopted stellar rotation period, whereas the MAP period is at 6.68 days in the sinusoid model}. The MAP eccentricity reaches { 0.12} in the 1-Planet model, which also suggests the Keplerian component is conforming to signals associated with stellar activity and rotation. Fit results for the three models are in Table~\ref{tab:results}.

We attempted fitting a GP model on RVs without conditioning from photometry. For each of these models we sampled $\lambda_e$, $\lambda_p$ and $P_{\rm GP}$ from prior distributions ${\mathcal LU}(1, 100)$, ${\mathcal LU}(0.1, 100)$, and ${\mathcal LU}(10, 50)$, respectively (${\mathcal LU}$ denotes the log-uniform distribution, with lower and upper bounds). Indeed, with appropriate amounts of data, observing cadence, and handling of model evidences, planets and GP noise parameters can be inferred from RVs alone \citep{Faria2016}. The GP-only model yields MAP values of ${\lambda_e = 30.44}$ days and ${P_{\rm GP} = 15.48}$ days. The GP + 1-Planet model returns ${\lambda_e = 46.43}$ days and ${P_{\rm GP} = 15.59}$ days. The returned planet has MAP orbital parameters ${K_s = 1.1}$ \ms, and ${P = 2.1}$ days. However, the GP-only model is favored at ${\Delta\ln {\mathcal Z} \simeq 1}$. While the inferred stellar rotation period is consistent with the photometry-derived rotation period, the spot evolution timescale is { closer to} twice the photometry value, probably due to the sparse sampling of the RV data.

Given that the model evidences are all within $|\Delta\ln {\mathcal Z}| < 2$ of each other, it is difficult to make decisive inferences through their comparison. The points that we would like to emphasize are that: (1) for this dataset, conditioning on high-cadence photometry provides important constraints on the stellar activity model that are otherwise difficult to infer from RVs alone, including a more accurate rotation period and spot evolution timescale; (2) consistent with previous analyses of other stars, the spot decay timescale tends to be longer than the stellar rotation period but within the same order of magnitude; and (3) the Bayesian evidence favors an activity model without planets. The GP's flexibility here is crucial since the activity signal quickly loses coherence, and spots follow various rotation periods depending on latitude. Both of these aspects are addressed by choosing a proper $\lambda_e$ and a single characteristic $P_{\rm GP}$, respectively.  In the following section we discuss the temporal sampling of our data, and how observing scheduling can improve the sensitivity of our analysis to short-period signals. In particular, additional high-cadence RVs are necessary to rule out some planets at the $K_s \approx { 3}$ \ms\ level.

\section{Importance of Cadence} \label{sec:cad}

\begin{figure}
    \centering
    \includegraphics[width=\linewidth]{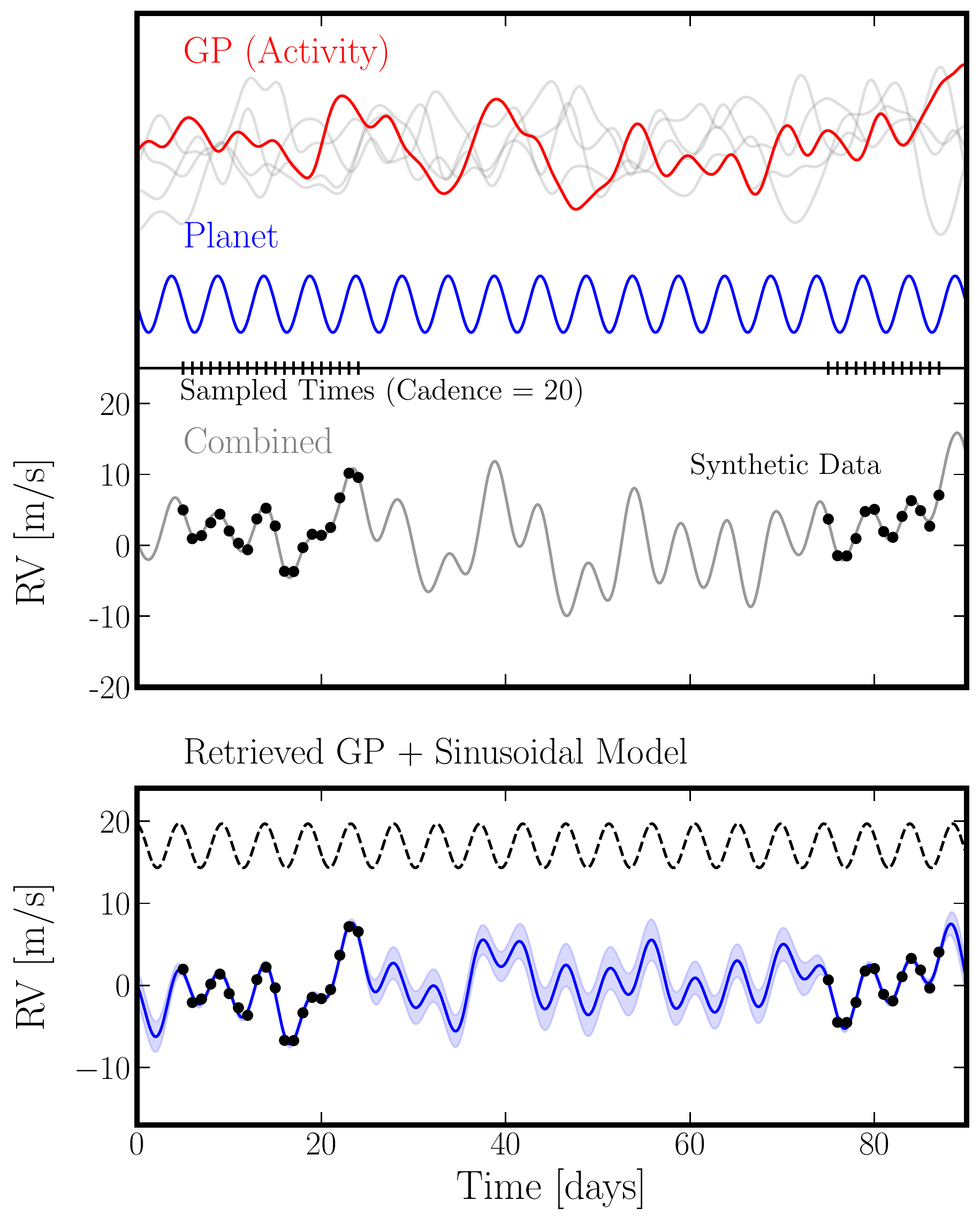}
    \caption{Illustration of the cadence analysis. The steps are as follows. (1) Generation of a synthetic RV dataset, consisting of a GP draw (activity model, red curve in top panel) and circular Keplerian (planet model, blue curve in top panel) (additional example GP draws are overplotted in gray). The two are added together and sampled at specific times based on the desired cadence. { There are 33 samples in total}. (2) Use of the GP framework to search for planet signals in the synthetic dataset (details in Section~\ref{sec:gp}). An example is shown in the bottom panel, including the GP + Sinusoidal mean and 1$\sigma$ confidence interval (blue line and shaded region) and the MAP Keplerian solution (dashed black line, offset for clarity). (3) Comparison between recovered orbital parameters and the true ones.}
    \label{fig:cadex}
\end{figure}

The RV dataset analyzed here presents a combination of fast cadence `bursts', as well as isolated data points. The high amplitude RV variations are well sampled in 2018 (top panel of Figure~\ref{fig:1p}), but { more poorly} sampled in 2019 and 2020 (middle and bottom panels). Cadence may be less important when Keplerians have higher amplitude than the stellar activity (e.g. hot Jupiters, or planets around quiet stars). However, high cadence is very useful when stellar activity dominates the time series, especially for detecting short-period planets. If RVs have cadence longer than spot lifetimes, then it is challenging for the GP model to learn the smoothness or periodicity of the activity signal.  Isolated data points (e.g. those acquired in 2020) provide little information in this sense. It is worth mentioning the recent discovery of a multi-planet system around GJ 887 \citep{Jeffers2020}. Confident detection was only made possible by a single, high-cadence observing season ($\sim 1$ exposure per clear night), even though there were nearly 20 years of existing data (on average $\sim11$ RVs per year). Another example is the case of Proxima Centauri b, in which one high-cadence observing season led to a higher detection significance than years of previous data \citep{Anglada2016}. Optimal cadence has been investigated before in the { contexts} of averaging out the activity contribution \citep{Dumusque2011} { and its relationship to orbital phase coverage \citep{Rajpaul2017}. Nightly coverage has also been compared to other ground and space-based schedules \citep{Hall2018}}. However, the relationship between observing cadence and GP modeling of activity has not previously been characterized. 

\subsection{An Injection \& Recovery Analysis }

To quantify the impact of cadence in our analysis, we perform a simple injection/recovery test as follows. First, we generate a synthetic RV time series by drawing a \texttt{george} GP specified by Equation~\ref{eqn:qpc}, characteristic of our HD~101501 observations. Hyperparameters { are set to the MAP values in \S\ref{subsec:photo}. The amplitude is set to 5.06 \ms, which is the MAP amplitude of the GP-Only fit in \S\ref{subsec:fitrv}}. Next, we add a Keplerian component with $K_s$ and $P$ sampled from a grid. Other Keplerian parameters are set to 0. We sample the continuous RV curve at { 33} epochs (described below) and add random noise drawn from a zero-mean Normal distribution with standard deviation { 40} \cms, which is also the associated uncertainty on each data point. Finally, we fit a GP + sinusoidal model as discussed in Section~\ref{sec:gp} with three GP hyperparameters fixed, as if they had been predetermined by photometry measurements. We then compare the { recovered} $K_s$ and $P$ to the actual injected signal. Each full run involves a pair ($K_s$, $P$) and new GP draw. { The sampler is run only once since we are not interested in precise uncertainties on the model evidence}. The procedure is illustrated in Figure~\ref{fig:cadex}.

The test described above is repeated for several toy observing cadences. We define an $N$-day cadence as observing for $N$ consecutive nights (separated by one sidereal day), followed by a long gap in time. The gap is drawn from a uniform random distribution between 50 and 80 days. This pattern repeats until the number of exposures totals { 33}, which was the number of nights HD~101501 was observed by EXPRES at the start of simulations (a couple of additional, recent nights were included in the RV analysis). Each timestamp is then perturbed by a uniformly random variable between $-4$ and $+4$ hours to simulate variations in observing scheduling. The $50-80$ day gap is about $3-5\times\lambda_e$, chosen to destroy coherence between consecutive bursts of exposures. A variable gap helps avoid unwanted sampling artifacts. We repeat the above test for 
$2$, $5$, $10$, $20$, and $33$-day cadences. For example, the $5$-day cadence involves timestamps for five consecutive nights, followed by a long gap in time. The pattern repeats six times, followed by { three} exposures such that the number of datapoints equals { 33}. The cadences roughly sample $1/8$, $1/4$, $1/2$, $1$, and $2$ stellar rotations. As a benchmark we also sample at the identical { timestamps of EXPRES data}.
These tests do not encompass sophisticated modeling of stellar noise or the wide variety of observing strategies and possible Keplerian signals (e.g. eccentric orbits, multiple planets). However, the assumptions made are reasonable for a goal of understanding how consecutive nights of observation relates to stellar activity inferences for stars like HD~101501. 

\begin{figure*}
    \centering
    \includegraphics[width=\linewidth]{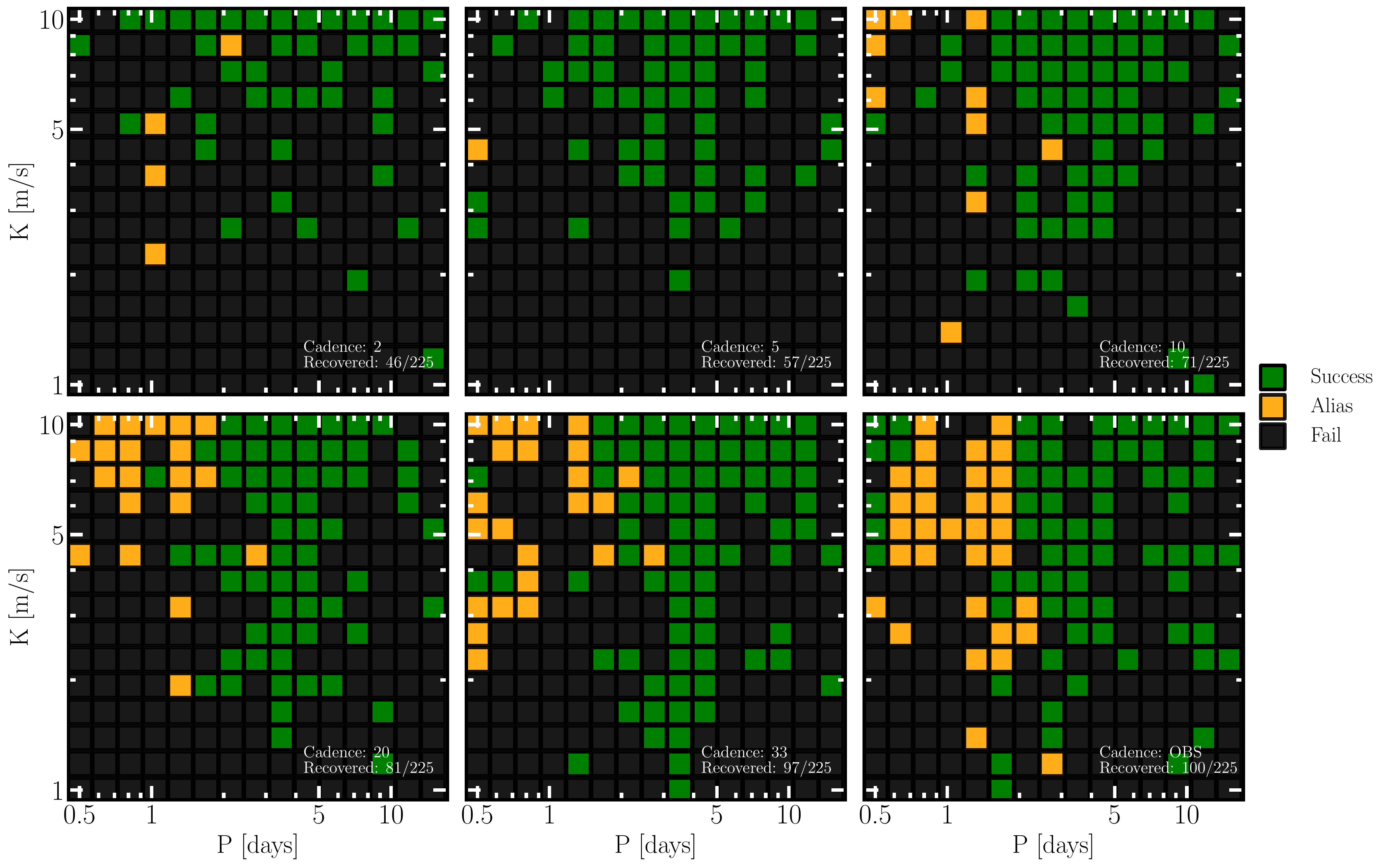}
    \caption{Results of the cadence analysis, in which the GP framework is applied to a synthetic RV time series, generated by a GP draw representing stellar activity and an added Keplerian. Each run involves a unique GP draw, and a combination of $K_s$ and $P$ drawn from a grid with log-uniform spacing. Green squares indicate successful recovery of the injected signal. Orange squares indicate recovery of an alias of the signal. Black squares indicate failure. The cadence stated in each plot (2, 5, 10, 20, { 33}) denotes the number of consecutive nights of observation before a long gap in time, and the total number of data points is { 33} in every run. The completion is generally higher for observing strategies using higher cadences. In particular, upwards of ten consecutive nights of observing are needed to detect some short period planets. For comparison, we also run the analysis on the epochs of our HD~101501 RV dataset.}
    \label{fig:sim}
\end{figure*}

\begin{figure}
    \centering
    \includegraphics[width=\linewidth]{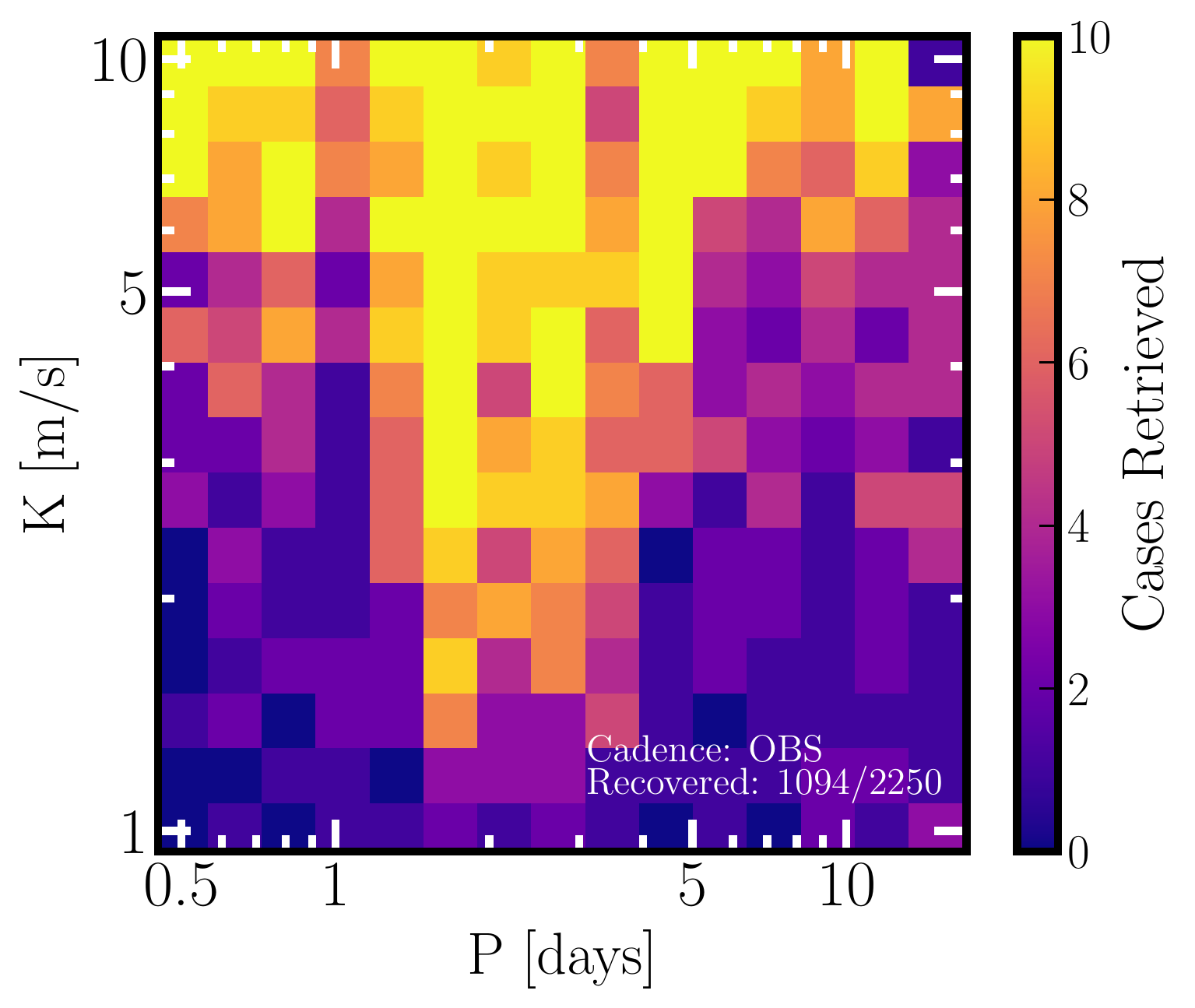}
    \caption{Sensitivity of the GP framework for identifying planets, given an observing cadence matching our HD~101501 dataset. The GP framework is applied to a synthetic RV time series, generated by a GP draw representing stellar activity and an added Keplerian. Each run involves a unique GP draw, and a combination of $K_s$ and $P$ drawn from a grid with log-uniform spacing. The RV curve is then sampled { at timestamps of EXPRES RV data}. The color scale represents the number of successful retrievals after repeating the injection/recovery 10 times.}
    \label{fig:simobs}
\end{figure}

\subsection{Cadence Simulation Results} \label{subsection:cadres}

The injected Keplerian is successfully recovered if the following criteria are met: first, { the recovered semi-amplitude must be greater than $3\sigma_K$, where $\sigma_K$ is the standard deviation of $K_s$ draws made by the nested sampler}; second, the fractional error on the period must be within $10\%$. We also check against $1.027$ day$^{-1}$ aliases of the recovered period \citep{Dawson2010} (1 day = 1.0027 sidereal day). Our injection + recovery analysis indicates that high observing cadence, consisting of several nights of consecutive observation, is necessary for the GP framework to identify certain classes of exoplanets orbiting active stars. Figure~\ref{fig:sim} depicts the recovery of planets for a given cadence (green denotes success, whereas orange denotes convergence on an alias). 

The 10, 20, and { 33}-day cadences have better completion across the parameter grid, especially for short periods. { The observed cadence also has comparatively good completion, mostly due to there being 76 total timestamps (multiple per night) instead of 33}. As expected, long-period/low-amplitude signals are difficult to recover for all cadences. We emphasize that this is not a full Monte-Carlo analysis, and that unsuccessful retrievals at $K_s > 5$ \ms\ and $P < 5$ days are likely a result of randomness in the GP draw and unfavorable sampling of the orbit. These high-amplitude Keplerians might be retrieved with a more careful analysis (e.g. selection of priors, more thorough posterior sampling, etc...). Planets with periods $\lesssim 1.5$ days are difficult to distinguish from their one-day aliases, but are nevertheless recovered by the GP framework. Most modern RV analyses are prone to occasionally identifying aliases over true signals \citep{Dumusque2017}. 

Importantly, the results show generally increasing completion with higher cadence. The 2-day cadence fails for nearly all cases except amplitudes $\gtrsim 5$ \ms, when the planet signal starts to rival the activity signal. The 5-day cadence exhibits improvements, and { 33}-day shows the greatest completion of $K_s$ and $P$ combinations. There are diminishing returns going from the 10-day to 20-day cadence, and the 20-day to { 33}-day cadence. For short-period/low-amplitude planets ($K_s \lesssim {3}$ \ms, $P \lesssim {5}$ days), if sampling covers multiple orbital periods within a couple spot lifetimes, then the high-frequency Keplerian can be distinguished from the low-frequency GP/activity component. In this case the actual stellar rotation period may be irrelevant and alternative kernels such as the squared-exponential may be suitable for modeling the activity \citep{Rajpaul2015}. The most challenging planets ($K_s \lesssim 2$ \ms) are retrieved best by the highest cadences (20-day and { 33}-day). In these cases, the GP { is capable of learning} repeated structure in the activity signal.

Given the above results, we would like to gauge whether the GP framework and EXPRES RVs of HD~101501 are actually sensitive to the full range of possible orbits specified by the priors. We perform a Monte-Carlo analysis in which we repeat the injection/recovery procedure, { sampled at the actual timestamps of exposures}, 10 times (i.e. we generate the bottom-right panel of Figure~\ref{fig:sim} an additional 10 times, each involving new GP draws). We group correct orbital periods and aliases together as `successful' recoveries. The results are shown in Figure~\ref{fig:simobs}. As expected the longest periods and smallest semi-amplitudes are most challenging to detect, as are periods at nearly 1-day. However, even for modest semi-amplitudes up to $\sim 3$ \ms, planets with periods less than a { couple} days are difficult to detect.

{ We explore two additional extensions to the above simulations. First, to what degree is detection reliant on the GP component? Second, how much easier is it to recover a planet with a known ephemeris (i.e. the planet transits)? These investigations involved modifying the underlying model to, in the first case, exclude a GP component and just fit a Keplerian, jitter term and offset term. For the second case we use the full model, but all Keplerian parameters are fixed to their true values except for semi-amplitude. The results of these tests are shown in the Appendix. Without a GP, the sampler has much greater difficulty identifying planets in the simulated RV datasets. Given the simulated cadences, a GP is necessary to detect most planets with $K_s < 5$. While the sampler does identify Keplerians at $P \sim 15$ days, the retrievals might be falsely converging on the stellar activity signal given the similar stellar rotation period. In the future, it would be useful to explore the efficacy of GPs on RV datasets when the planet period is near the stellar rotation period. On the other hand, when the ephemeris is known and fixed {\it a priori}, we see a dramatic improvement in recovery when a GP is used. Robust detections are difficult for only the lowest semi-amplitudes and longest periods, limited by factors such as phase coverage, time baseline, and measurement uncertainty.}

\section{Discussion and Conclusions} \label{section:disc}

We discuss some additional aspects of the data which were not thoroughly explored in the above GP analysis. For example, we have thus far neglected activity indicators (BIS, FWHM, etc.), and restricted attention to the high-fidelity photometry. Indeed, when simultaneous photometry is unavailable, indicators can become important in providing additional constraints on spot presence and evolution. We also analyze the photometry in greater detail by investigating each season's characteristics and performing light-curve inversion.

\subsection{Activity Indicators}

\begin{figure*}
    \centering
    \includegraphics[width=\linewidth]{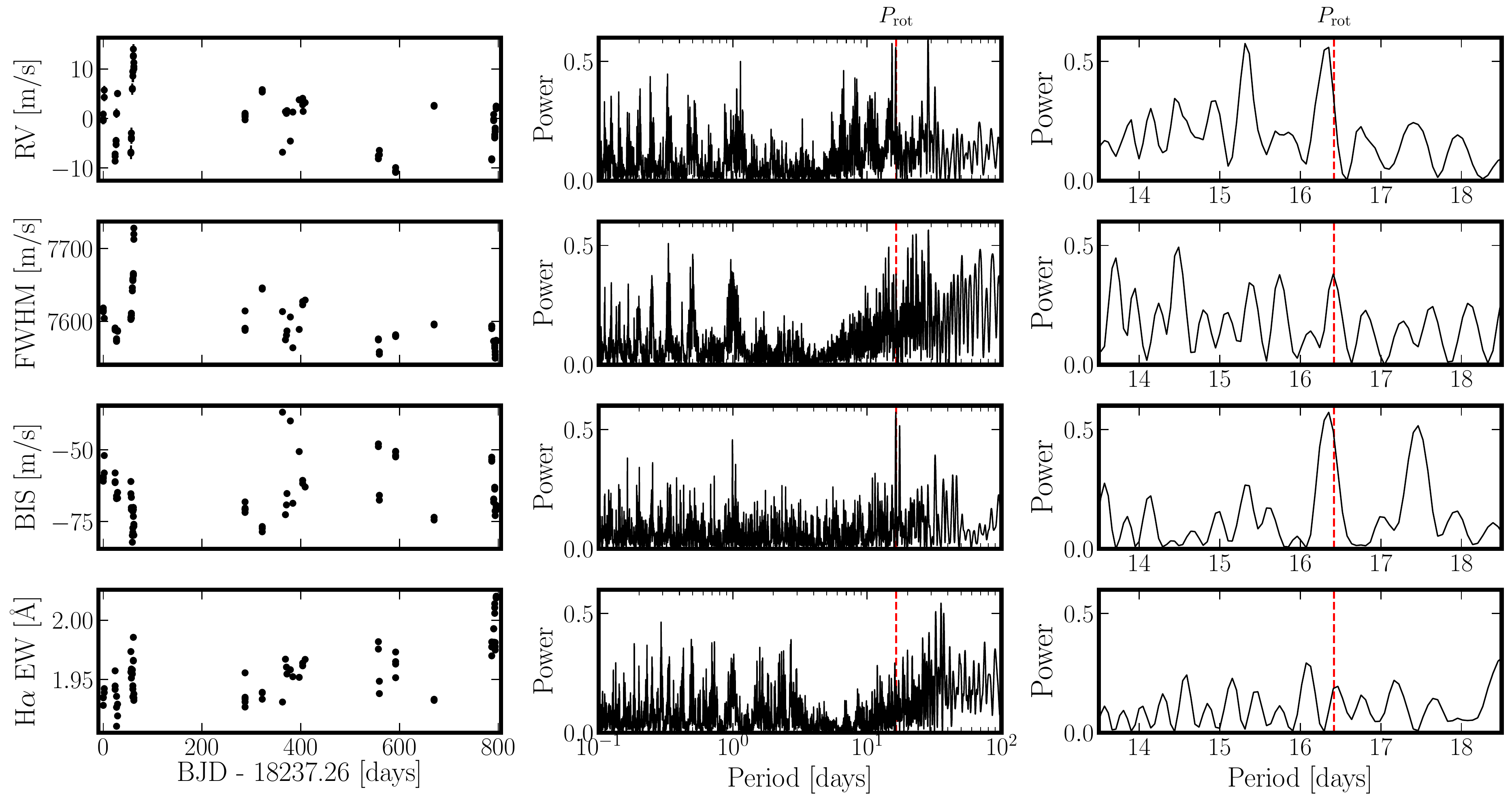}
    \caption{RV and activity indicator time series (left panels), and their periodograms (middle panels). The red dashed line denotes the stellar rotation period, as derived in this study. The periodograms are shown again in a cropped region around the stellar rotation period (right panels)}
    \label{fig:ind}
\end{figure*}

Several approaches directly incorporate indicators into the GP framework. Indicators may be modeled jointly with RVs as linear combinations of a single, latent GP and its derivative \citep{Rajpaul2015, Jones2017, Gilbertson2020}. Another method involves regression on RVs and an indicator time series where both GPs share certain hyperparameters \citep{Mascareno2020}; or, one can train a GP on an indicator time series and use the results as an initial guess for a subsequent RV analysis \citep{Dumusque2017}. 

In Figure~\ref{fig:ind} we show the RV time series along with CCF FWHM, CCF BIS, and H$\alpha$ EW, as well as their Generalized Lomb-Scargle periodograms \citep{Scargle1982, Zechmeister2009}. The RV and BIS time series have power at the stellar rotation period, with BIS more pronounced. BIS shows a double-peaked structure, split between roughly the GP-derived rotation period and another peak near $\sim 17.2$ days. This pattern is likely due to differential rotation, where long-lived spots at different latitudes exhibit different rotation periods. We attempted joint-modeling of RVs and BIS with separate Gaussian Processes that share $\lambda_e$, $\lambda_p$, and $P_{\rm GP}$. Their likelihood functions were summed together. We sampled the BIS GP amplitude, mean, and jitter from similar distributions as used for the RVs. The 0-Planet model is favored over the 1-Planet model, with ${\ln {\mathcal Z}_0 = -363.98 \pm 0.07}$ and ${\ln {\mathcal Z}_1 = -364.92 \pm 0.13}$, respectively. The corresponding MAP values of $P_{\rm GP}$ are ${15.87}$ days and ${15.65}$ days, respectively, { similar to the rotation period dervied from RVs alone}. The sampler converges to ${\lambda_e = 34.01}$ days and ${\lambda_e = 27.47}$ days for the two models, respectively. The APT light curve changes significantly between rotations, so the actual spot evolution is likely much faster than these estimates.

We { make} a few comments on the periodograms, which may simply be spurious features of the dataset. The RV periodogram shows highest power at $\sim 28$ days; however, none of { the} GP models with planets converged to this orbital period in additional trials where we extended the orbital period prior to 40 days. Rather, they still tended toward the other peak at $15.3$ days. Since the indicators do not have significant power at $15.3$ days, a Keplerian might be responsible. However, we reiterate that the Bayesian evidence disfavors a planet in all of our fits (light curve GP conditioning, BIS joint fitting, and freely fitting RVs), and $15.3$ days is close to the stellar rotation period.

\subsection{Photometric Variability}

Our RV analysis warranted use of the most recent seasons of photometry. However, the remarkable 28-season baseline offers an opportunity to obtain more precise constraints on stellar rotation and activity. We again turn to GP regression, which has seen frequent applications toward inferring stellar properties and searching for transits in photometry \citep{Vanderburg2015, Angus2018, Barros2020}. We use a \texttt{celerite} GP with covariance function given by Equation~\ref{eqn:k2} and sample hyperparameters $\{B, C, L, P_{\rm GP}\}$. We find a periodic timescale $P_{\rm GP} = {16.45_{-0.51}^{+0.60}}$ days, and an evolutionary timescale of $L = {15.82_{-3.07}^{+9.04}}$ days (uncertainties denote $16^{\rm th}$ and $84^{\rm th}$ percentiles around the median). The MAP $P_{\rm GP}$ is ${16.42}$ days, which we take as our best estimate of the rotation period $P_{\rm rot}$. Technically, this is probably not the equatorial rotation period. However, it is the period that best describes the data, and it might be influenced by the typical latitudes of spots. The GP with MAP hyperparameters is shown in the bottom panel of Figure~\ref{fig:aptfull}. The evolutionary timescale $L$ is consistent within $1\sigma$ of $\lambda_e$ from our earlier analysis where we fit a \texttt{george} GP to the recent photometry. The periodic timescale $P_{\rm GP}$ also matches between the two fits. { Formally, $P_{\rm GP}$ has a different definition in Equation~\ref{eqn:qpc} than it does in Equation~\ref{eqn:k2}, but it has similar meaning and influence on GP behavior}. Their similarity gives some assurance that the stellar activity signal in the most recent three seasons shares similar characteristics with those of the whole baseline. The closeness of $L$ and $P_{\rm rot}$ indicates that the spot distribution changes significantly between consecutive rotations, which also makes the light curve less coherent.

Some of our photometry seasons (e.g. 2001) have very coherent oscillations, while others (e.g. 1998) appear more random. We investigate inter-seasonal variations by fitting a \texttt{celerite} GP to each season individually (Figures~\ref{fig:aptseason1} $\&$ \ref{fig:aptseason2}). The GP provides, in theory, a more reliable estimate of the rotation period than the maximum power of the periodogram. The periodogram { is based on a single sinusoid model and most clearly identifies a signal when the phase, amplitude and period are constant. The quasiperiodic GP can accommodate signals that exhibit small departures from an overall phase, amplitude and period, for example due to appearance and disappearance of spots.} However, the robustness of the GP is tied to the quantity and cadence of the data (lacking in 2017, for example). Also, if multiple modes are present within a given season, the GP learns a value that maximizes the likelihood, which might not actually be representative of any single mode. In ${10}$ of 28 seasons the best-fit GP has $P_{\rm GP} < 16$ days, reaching as low as ${12-14}$ days. It is below 17 days in all cases. The variability in periodic timescales is most likely a sign of differential rotation. In a previous study, \citet{Mittag2017} analyze periodograms of Ca II H\&K and Ca II IRT line strengths and find power at multiple periods, which they also attribute to differential rotation.  We perform a bootstrap by sampling the 28 values of $P_{\rm GP}$ with replacement 10,000 times, recording the rotational shear $\alpha = \Delta P/P$ at each iteration. The difference between maximum and minimum periods, $\Delta P$, assumes that the maximum corresponds to rotation at the poles and the minimum at the equator. We find $\alpha = {0.45_{-0.19}^{+0.03}}$, which is similar to the solar rotational shear of $\sim 0.4$ \citep{Snodgrass1990}.

\subsection{Light-curve Inversion}\label{subsec:LI}

\begin{figure*}
    \centering
    \includegraphics[width=0.72\linewidth]{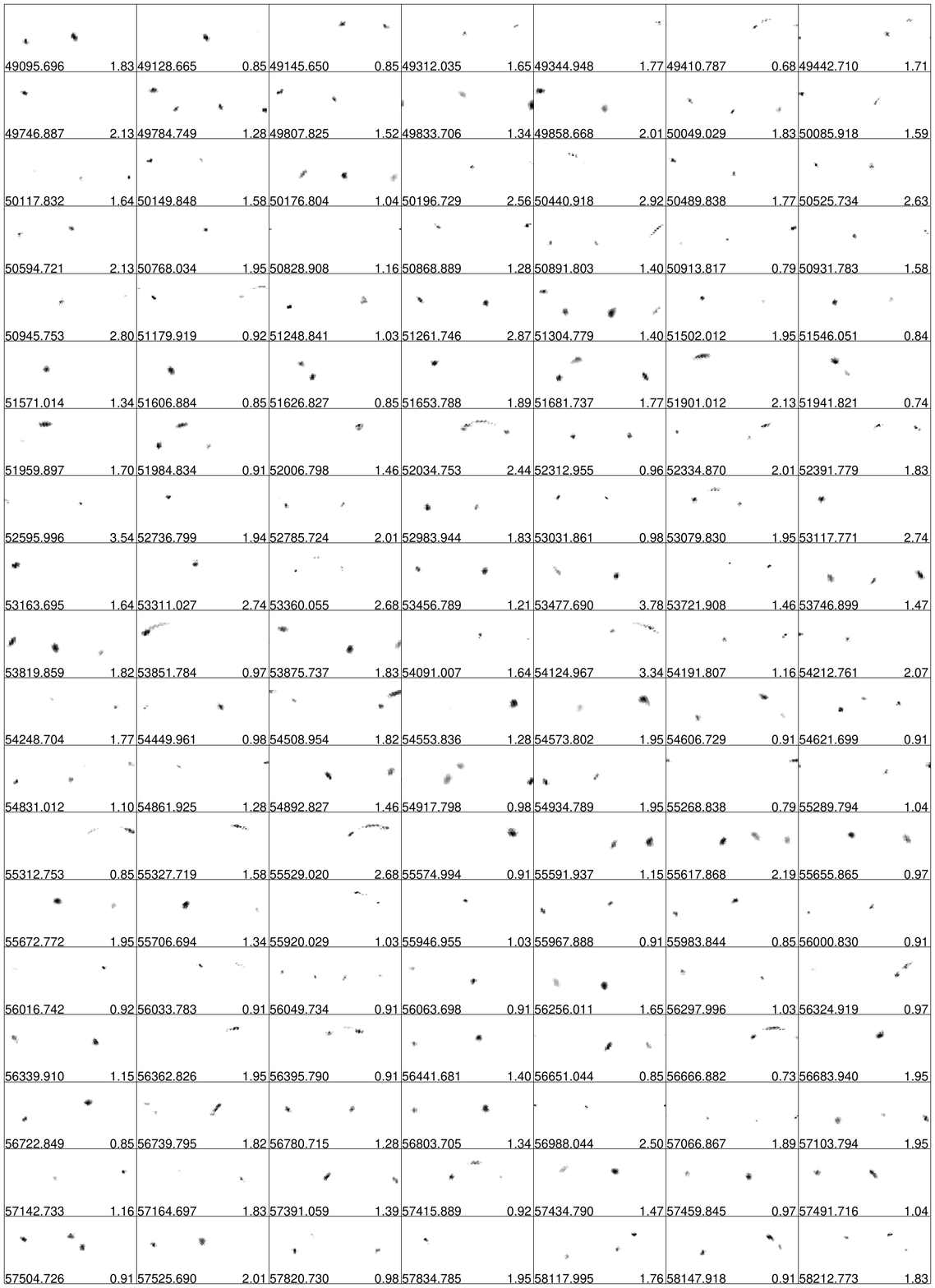}
    \vspace{-.5cm}
    \caption{Pseudo-Mercator surfaces of HD~101501 reconstructed using LI.  The date of the earliest data point used in the light curve are included in the lower left corner of each plot (JD - 2400000.5).  Each of the surfaces included here use data that were collected before the EXPRES data.  The number of rotations used in each inversion is included in the lower right corner of each plot. Each plot shows all stellar longitudes horizontally and all stellar latitudes vertically.  The center of the star, as visible to the observer, at phase 0.25 is located at $0^\circ$ longitude, and at the left edge of the surfaces here.  The star rotates over time with the longitudes \emph{decreasing}.}
    \label{fig:LI1}
\end{figure*}

\begin{figure}
    \centering
    \includegraphics[width=\linewidth,trim={2.0cm 6.0cm 5.0cm 0.5cm},clip]{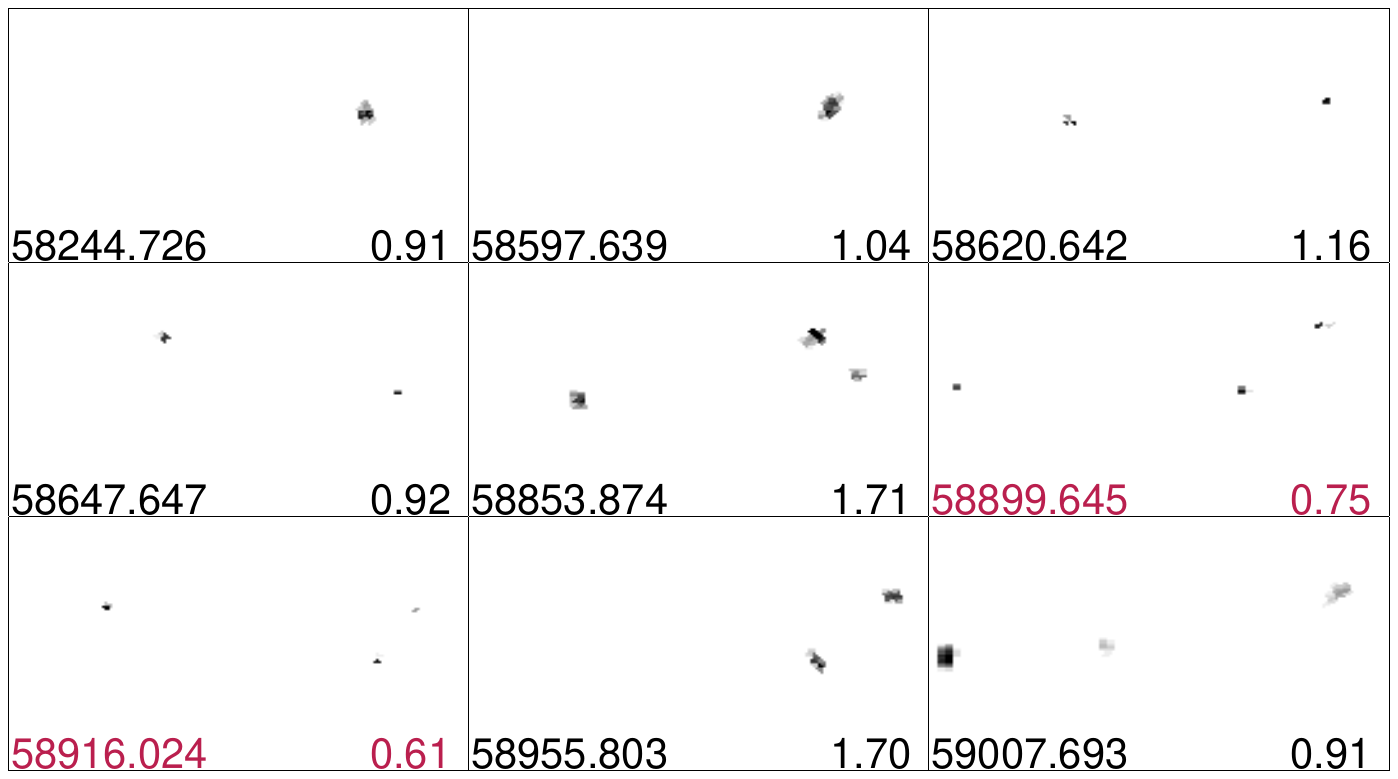}
    \vspace{-8cm}
    \caption{Pseudo-Mercator surfaces of HD~101501 reconstructed using LI, as in Figure \ref{fig:LI1}.  The surfaces here use photometric data obtained during the epochs of EXPRES data.  The two surfaces with MJD and number of rotations used in the inversion written in red are reconstructed from \emph{TESS} data.}
    \label{fig:LI2}
\end{figure}

In order to better understand the evolution of the starspots on the surface of HD~101501, we applied a light-curve inversion algorithm to reconstruct the stellar surfaces.  The algorithm Light-curve Inversion \citep[LI;][]{Harmon2000} uses a modified-Tikhonov regularization and stellar parameters to converge on a solution and makes no \emph{a priori} assumptions about starspot shape, number, or size.  

For the APT light curve, we divided the light curve into portions lasting approximately 1 rotation.  When necessary to provide more phase coverage, the portion of the light curve used for inversion lasted more than one rotation (this is noted in the lower right corner in Figures~\ref{fig:LI1} and~\ref{fig:LI2}).  In very few cases, data points were used for overlapping portions of the light curve.  For the \emph{TESS} light curve, there is a gap over four-days long in the middle of the sector of observation (see Figure \ref{fig:TESSLC}.  We divided the light curve during this time into two rotations and binned the data into bins 0.01 in phase.  Both rotations {observed with \emph{TESS}} have incomplete phase coverage.  

We assigned \teff $ = 5500$~K and a spot temperature  T$_\mathrm{spot} = 4500$~K, based on Figure 7 of \citet{Berdyugina2005}.  Combining those temperatures and the appropriate response function for the filters, yields a spot-to-photosphere ratio of 0.3452 for $V$-band and 0.4530 for the \emph{TESS} observations.  In addition to this value, LI also uses limb-darkening coefficients as input.  For the $V$-band APT inversions, quadratic limb-darkening coefficients are used \citep[0.5955 and 0.1488,][]{Claret2013}.  For the \emph{TESS} inversions, quadratic limb-darkening coefficients are also used \citep[0.3876 and 0.2036,][]{Claret2018}.  Each inversion used a unique root-mean-squared (rms) error in order to balance the inversion algorithm between over-fitting and over-smoothing the data (the average rms for the APT light curves was 0.0013 and both \emph{TESS} light curves used an rms of 0.0003).  
A stellar inclination of $i = 52^\circ$ was found from the $v \sin i$ given by \citet{Brewer2016}, the stellar diameter \citep{Bonneau2006}, and the parallax \citep{Gaia2018}.  

The reconstructed surfaces can be found in Figures \ref{fig:LI1} and \ref{fig:LI2}.  {The reconstructed dark regions on the stellar surface are not necessarily individual starspots, but may be unresolved groups of spots.  We cannot differentiate between these possibilities, and we refer to the dark regions as starspots.} The starspots of HD~101501  {behave similarly to} sunspot {groups} in that they typically evolve on the same timescale as a rotation.     On some occasions, the same spot appears to be present for more than one rotation.  The latitudes at which the starspots appear are only weakly constrained by the light curve and the limb-darkening coefficients used.  Because of this and the short starspot lifetimes, we do not further investigate differential rotation for this star.  {The variations in the photometric data and the resultant surface reconstructions indicate that a}t all points of observation, {there is evidence of} starspots on the surface.  Furthermore, the surfaces that result from the inversion of \emph{TESS} light curves show similar structures, but are not a direct comparison to the $V$-band reconstructions because the bandpasses are very different.  The \emph{TESS} filter covers a much larger wavelength range, and the contrast between cool starspots and the photosphere is more dramatic at shorter wavelengths.  

\subsection{Conclusions} \label{section:con}
The Sun-like star HD~101501 presents an interesting case study for understanding stellar activity signatures in radial velocities and photometry. We present new high-precision radial velocities of HD~101501 along with simultaneous ground-based photometry with a baseline of 28 years. Several weeks of photometry from \emph{TESS} are also analyzed. We summarize our findings as follows:

\begin{itemize}
    \item A Gaussian Process framework is used to model both the photometry as well as the correlated noise in the RV time series. HD~101501 represents a case study for { the} GP framework that may be applied to RVs of other EXPRES targets. The radial velocities, which exhibit variations at a level of $\sim 10$ \ms, are best explained by an activity-only model. The Bayesian evidence disfavors the presence of a Keplerian signal in our data. 
    \item Through a simple injection and recovery analysis, we explore the space of orbital parameters to which the GP framework is sensitive. We test different cadences in order to understand how observing strategy impacts our ability to detect planets. The lowest cadences, in which exposures are largely spaced in time, contribute very little to the GP retrieval. Higher cadences, in which the star is observed for many consecutive nights, assist the GP framework in separating short-period orbits from the stellar activity signal. These results refine our observing plans and offer important guidance for RV observations of active stars.
    \item GPs place tight constraints on the stellar rotation period associated with spots, at $P_{\rm rot} \sim {16.4}$ days. We use GPs to analyze periodicity in individual photometric observing seasons. The variability from season-to-season suggests a rotational shear of $\sim {0.45}$ and an equatorial rotation period of $\sim13$ days. 
    \item Reconstructed stellar surfaces show the persistent presence of starspots on the surface of HD~101501 at all times.  While starspots are always present, they are observed to change significantly between rotations making it impossible to trace their evolution over many rotations in these data.  
\end{itemize}

Detecting exoplanets around active stars remains a significant challenge. Correlated noise models show great promise for mitigating activity in RVs, especially when combined with simultaneous photometry. GPs have had great success in modeling stellar activity in HARPS, HARPS-N, CARMENES and ESPRESSO RVs, and this analysis of HD~101501 represents the first application of GPs to EXPRES RVs. The high-precision of current spectrographs, optimized observing strategy, and new RV extraction techniques will push exoplanet detection limits in the near future.

\acknowledgments
This work used data from the EXtreme PREcision Spectrograph (EXPRES) that was designed and commissioned at Yale with financial support by the U.S. National Science Foundation under MRI-1429365 and ATI-1509436 (PI D. Fischer). We gratefully acknowledge support for telescope time using EXPRES at the LDT from the Heising Simons Foundation and an anonymous Yale donor. We acknowledge critical support for investigation of photospheric noise in RV data from the NSF AST-1616086 and NASA 80NSSC18K0443.  R.M.R. acknowledges support from the YCAA Prize Postdoctoral Fellowship. G.W.H. acknowledges long-term support from NASA, NSF, Tennessee State University, and the State of Tennessee through its Centers of Excellence program. { We also thank the anonymous referee whose comments improved the cadence analysis and overall clarity of the manuscript.}

\bibliographystyle{aasjournal} 
\bibliography{main} 

\appendix

\pagebreak

\renewcommand{\arraystretch}{1.0}
\begin{table*}
\setlength{\tabcolsep}{8pt}
{
\begin{tabular}{ l l r r r r r r }
\hline
  Parameter & Units & GP-Only & & GP + 1-Planet & & GP + Sinusoid & \\
\hline
\hline
  $a$                & \ms  & $ 5.94_{-1.43}^{+11.51}$& (5.06)      & $ 5.38_{-0.70}^{+6.91}      $ & (4.49)     & $ 5.47_{-0.80}^{+7.98}      $ & (5.04)   \\
  $\lambda_e$        & days & $ 18.74^* $             & $^*$        & $^*$                          & $^*$       & $ ^*                        $ & $^*$  \\
  $P_{\rm GP}$       & days & $ 16.28^* $             & $^*$        & $^*$                          & $^*$       & $ ^*                        $ & $^*$  \\
  $\lambda_p$        & -    & $ 0.76^* $              & $^*$        & $^*$                          & $^*$       & $ ^*                        $ & $^*$ \\
  $s$                & \ms  & $0.38_{-0.20}^{+1.42}$  & (0.27)      & $ 0.30_{-0.10}^{+0.81}      $ & (0.16)     & $ 0.31_{-0.11}^{+0.87}      $ & (0.31)  \\
  $\gamma$           & \ms  & $-0.41_{-4.89}^{+5.27}$ & (-0.41)     & $ -0.42_{-2.84}^{+2.86}     $ & (-0.70)    & $ -0.44_{-3.11}^{+3.22}     $ & (-0.60) \\
  $K_s$              & \ms  & -                       & -           & $ 0.63_{-0.43}^{+1.19}      $ & (4.02)     & $ 0.59_{-0.39}^{+1.18}      $ & (2.53)  \\
  $\phi_0$           & rad. & -                       & -           & $ 3.11_{-1.87}^{+1.93}      $ & (3.59)     & $ 3.03_{-1.85}^{+2.03}      $ & (3.87)  \\
  $P$                & days & -                       & -           & $ 3.95_{-2.64}^{+13.03}     $ & (15.24)    & $ 4.20_{-3.00}^{+13.63}     $ & (6.68)   \\
  $\omega$           & rad. & -                       & -           & $ 3.06_{-1.96}^{+2.01}      $ & (3.90)     & $0^*                        $ & $^*$  \\
  $e$                & -    & -                       & -           & $ 0.03_{-0.03}^{+0.27}      $ & (0.12)     & $0^*                        $ & $^*$  \\
\hline
$\ln {\mathcal Z}$ & - & $-149.26 \pm 0.15$ & & $-149.72 \pm 0.06$ & & $-149.75 \pm 0.18$ & \\
$\ln {\mathcal L_{\rm MAP}}$ & - & $-139.80$ & & $-134.09$ & & $-135.81$ & \\
rms                & \cms & 0.45 & & 0.46 & & 0.45 & \\
\hline
\end{tabular}
}
  \caption{Results of our \texttt{george} GP retrieval. The three models correspond to GP-Only (no planet), GP + 1-Planet, and GP + Sinusoid (one planet, restricted $w$ and $e$). Columns contain the { median} of the marginalized distribution of each sampled parameter, and uncertainties correspond to $16^{\rm th}$ and $84^{\rm th}$ percentiles. Values in parentheses `` () " denote the {\it maximum a posteriori} (MAP). Asterisks `` * " denote fixed values, either from conditioning GP hyperparameters on the photometry data, or from restricting the Keplerian to a circular orbit. Missing values `` - " denote the parameter is not used in the model. The bottom rows contain the log-evidences returned by the nested sampler, { the log-likelihood of the MAP vector, and the rms of the residuals}. { Uncertainties on model evidences correspond to the standard deviation after five separate runs of the sampler.}}
  \label{tab:results}
\end{table*}

\begin{figure*}
    \centering
    \includegraphics[width=0.8\linewidth]{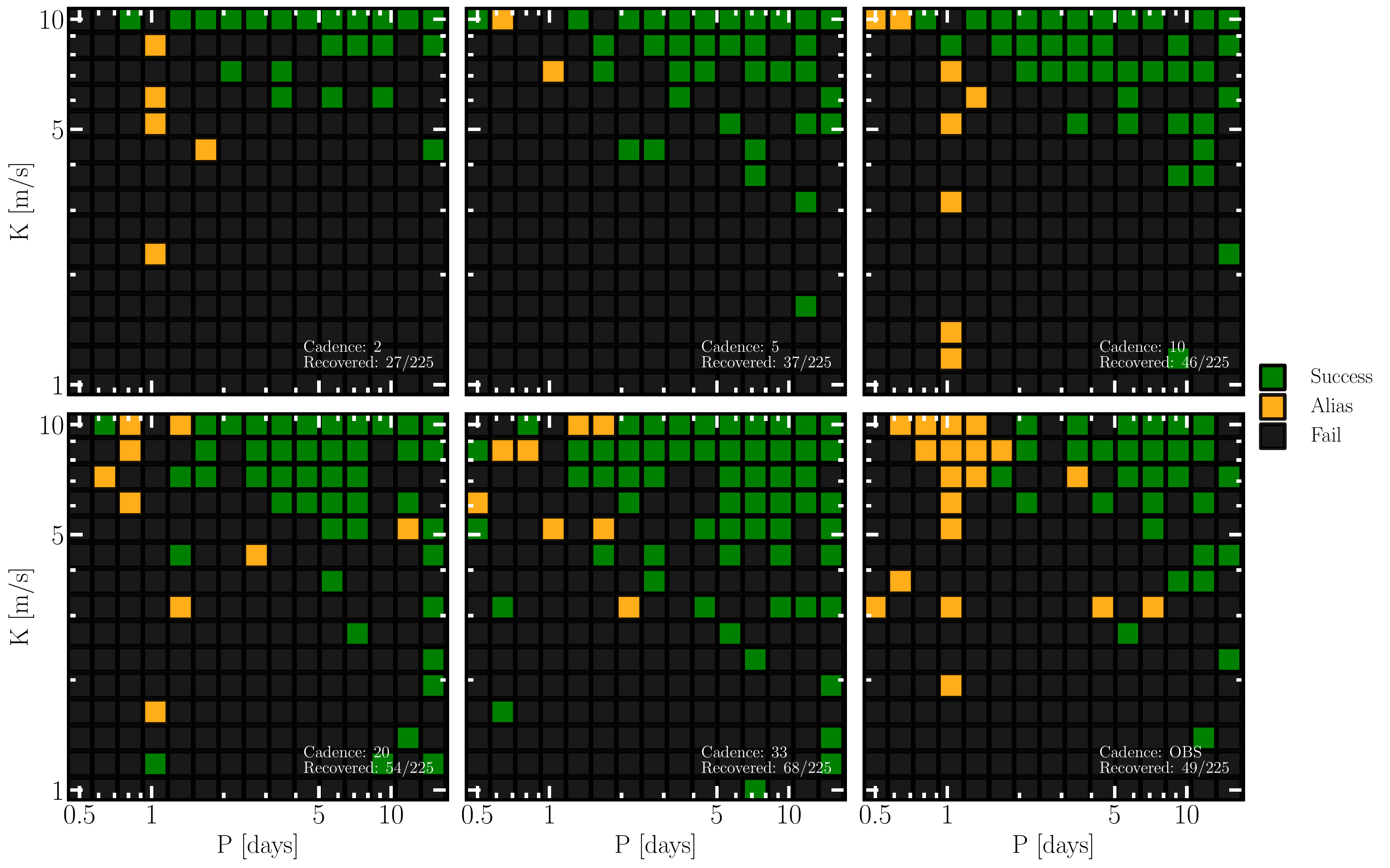}
    \caption{Same as Figure \ref{fig:sim}, except the model did not include a GP component. It included an offset term, jitter term, and sinuoid component. These simulations show significantly less completion compared to retrieval with the activity GP component.}
    \label{fig:simnogp}
\end{figure*}

\begin{figure*}
    \centering
    \includegraphics[width=0.8\linewidth]{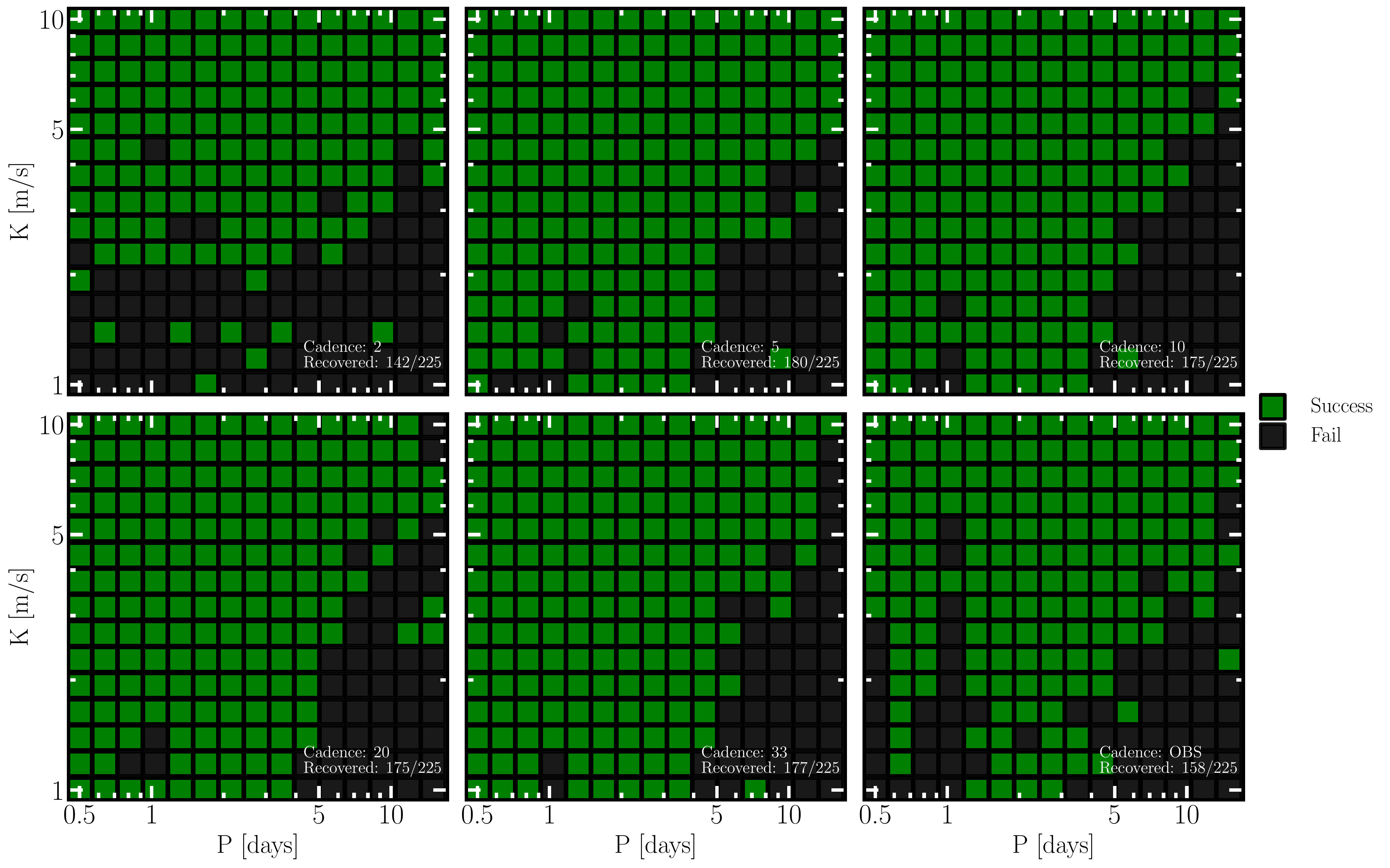}
    \caption{Same as Figure \ref{fig:sim}, except the model had period and phase fixed to true values. Such a model may be appropriate for real data when an ephemeris is known from primary transit.}
    \label{fig:simpfix}
\end{figure*}

\begin{figure*}
    \centering
    \includegraphics[width=\linewidth]{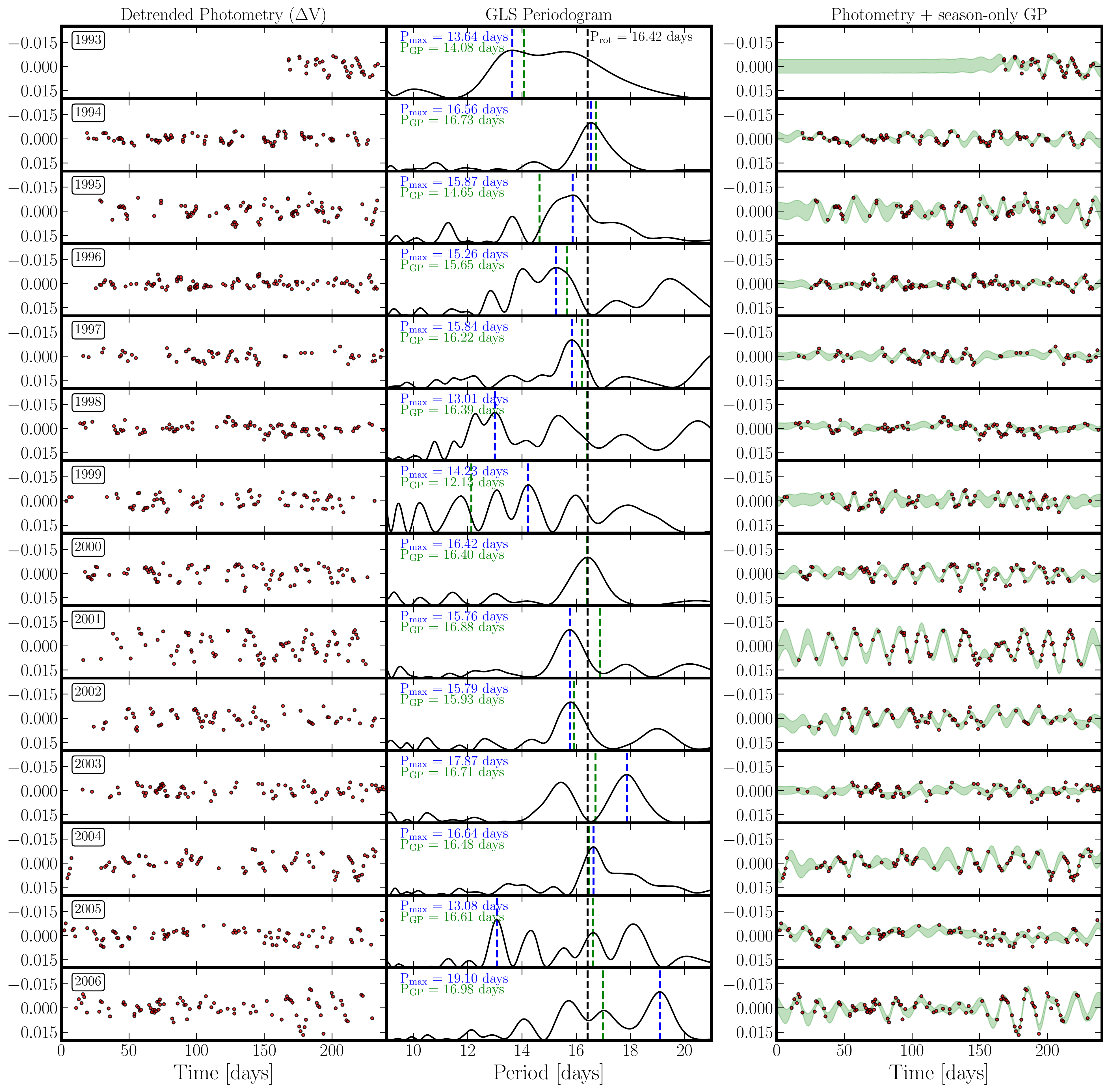}
    \caption{Analysis of individual APT photometry seasons (1993-2006). The left side of the figure shows the photometry data along with their GLS Periodograms. The three periods marked are: 1) maxmimum power in the periodogram (blue); 2) the periodic timescale hyperparameter of a \texttt{celerite} GP, fit to that season of data (green); and 3) the rotation period of the star (black). The right panels show the \texttt{celerite} GP fit to each season (1$\sigma$ confidence interval).}
    \label{fig:aptseason1}
\end{figure*}

\begin{figure*}
    \centering
    \includegraphics[width=\linewidth]{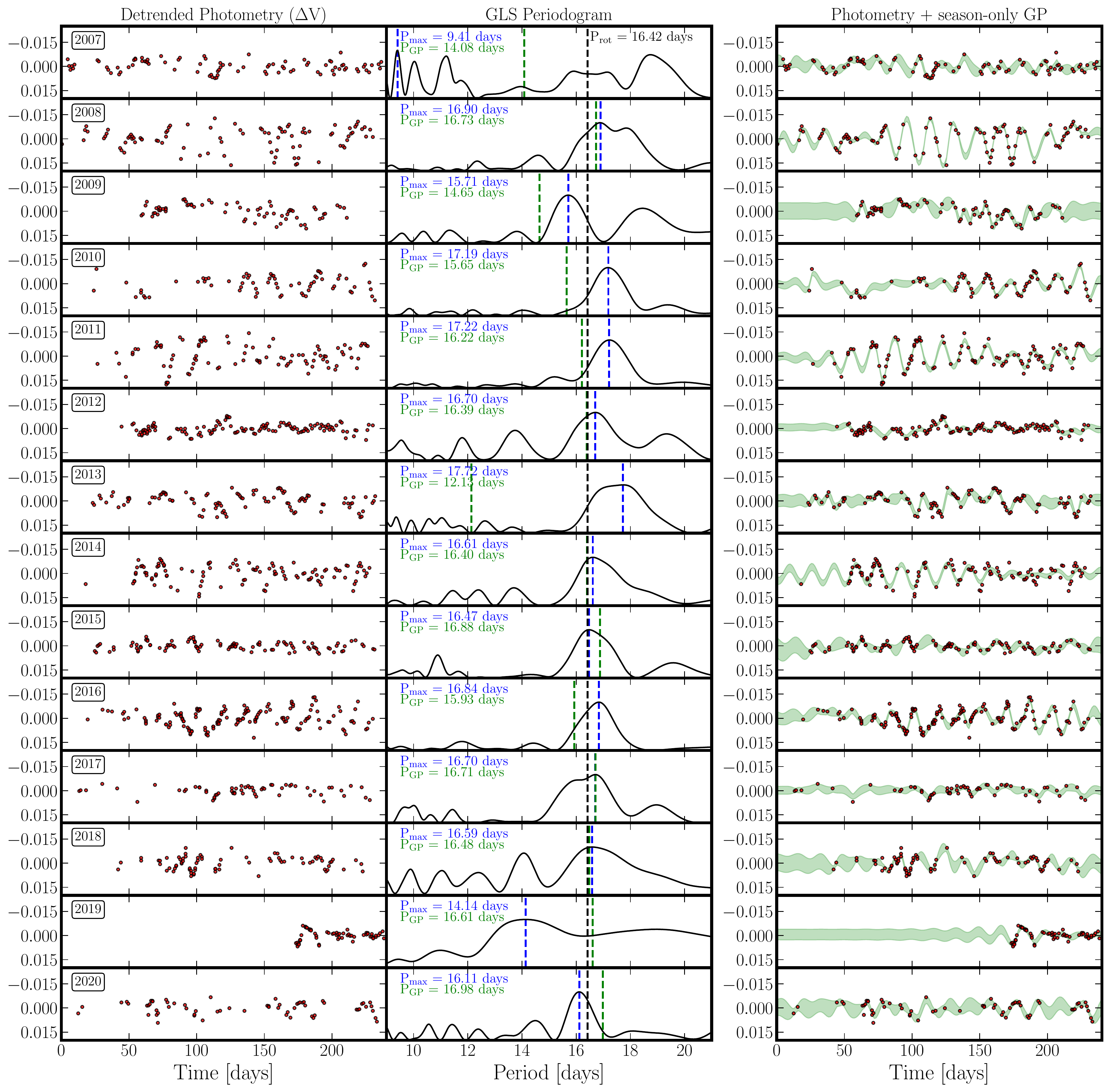}
    \caption{Analysis of individual APT photometry seasons (2007-2020). The left side of the figure shows the photometry data along with their GLS Periodograms. The three periods marked are: 1) maxmimum power in the periodogram (blue); 2) the periodic timescale hyperparameter of a \texttt{celerite} GP, fit to that season of data (green); and 3) the rotation period of the star (black). The right panels show the \texttt{celerite} GP fit to each season (1$\sigma$ confidence interval).}
    \label{fig:aptseason2}
\end{figure*}

\setlength{\tabcolsep}{2pt}
\def\arraystretch{1.0}
\begin{table*}
\begin{tabular}{lr}
\begin{minipage}{.5\linewidth}
\begin{tabular}{cccccc}
\hline
\hline
Time & Vel. & Unc. & {H$\alpha$ EW} & {BIS} & {CCF FWHM} \\
{(JD$-$2440000)} & {(m/s)} & {(m/s)} & {(\AA)} & {(m/s)} & {(m/s)} \\
\hline
 18237.2600 &  0.866 & 0.735 & 1.9347 & -59.4 & 7613.3 \\
 18237.2650 & -0.183 & 0.809 & 1.9283 & -60.9 & 7618.3 \\
 18237.2690 & -0.406 & 0.807 & 1.9351 & -59.9 & 7615.2 \\
 18239.2394 &  4.284 & 0.838 & 1.9391 & -51.9 & 7604.1 \\
 18239.2441 &  5.736 & 0.821 & 1.9424 & -58.0 & 7604.1 \\
 18261.2177 & -7.176 & 0.697 & 1.9574 & -61.4 & 7589.8 \\
 18261.2222 & -8.618 & 0.686 & 1.9446 & -61.0 & 7588.0 \\
 18261.2268 & -7.653 & 0.691 & 1.9417 & -58.0 & 7590.9 \\
 18263.2070 & -4.434 & 0.715 &      - &   -   &     -  \\
 18263.2117 & -5.267 & 0.762 &      - &   -   &     -  \\
 18263.2193 & -4.526 & 0.796 &      - &   -   &     -  \\
 18264.1671 &  0.977 & 0.847 & 1.9359 & -67.0 & 7576.1 \\
 18264.1717 &  1.105 & 0.899 & 1.9267 & -66.0 & 7573.8 \\
 18264.1762 &  0.932 & 0.813 & 1.9107 & -66.9 & 7572.5 \\
 18266.1937 &  5.011 & 0.642 & 1.9293 & -64.8 & 7587.4 \\
 18266.2010 &  5.070 & 0.650 & 1.9194 & -66.8 & 7586.3 \\
 18293.1759 & -7.014 & 1.204 & 1.9564 & -65.3 & 7602.6 \\
 18293.1817 & -6.800 & 1.374 & 1.9736 & -61.0 & 7605.4 \\
 18294.1745 & -4.139 & 0.959 & 1.9589 & -66.5 & 7608.3 \\
 18294.1789 & -3.846 & 0.911 & 1.9558 & -70.1 & 7604.4 \\
 18294.1833 & -2.942 & 1.071 & 1.9513 & -70.8 & 7611.1 \\
 18296.1927 &  5.893 & 1.136 & 1.9582 & -82.2 & 7641.9 \\
 18296.1974 &  6.064 & 1.029 & 1.9547 & -79.8 & 7646.1 \\
 18297.1717 &  9.454 & 1.068 & 1.9420 & -78.8 & 7658.6 \\
 18297.1763 &  9.517 & 1.226 & 1.9448 & -71.4 & 7655.9 \\
 18297.1809 &  8.567 & 1.245 & 1.9355 & -77.2 & 7656.5 \\
 18298.1707 & 12.705 & 1.004 & 1.9662 & -70.8 & 7665.6 \\
 18298.1752 & 12.527 & 0.998 & 1.9856 & -73.2 & 7664.6 \\
 18298.1798 & 14.018 & 1.012 & 1.9657 & -70.0 & 7662.8 \\
 18299.1760 & 10.011 & 0.912 & 1.9325 & -76.6 & 7727.9 \\
 18299.1806 & 11.297 & 0.937 & 1.9346 & -79.7 & 7719.7 \\
 18299.1852 & 10.481 & 0.967 & 1.9379 & -75.9 & 7712.5 \\
 18524.4706 &  0.594 & 0.324 & 1.9312 & -71.8 & 7589.8 \\
 18524.4840 &  0.340 & 0.756 & 1.9557 & -68.1 & 7614.2 \\
 18524.4961 &  1.055 & 0.303 & 1.9337 & -70.7 & 7590.5 \\
 18524.5017 &  0.873 & 0.313 & 1.9269 & -70.8 & 7590.1 \\
 18524.5078 & -0.264 & 0.330 & 1.9351 & -70.3 & 7587.6 \\
 18559.2641 &  5.351 & 0.325 & 1.9335 & -78.6 & 7644.2 \\
\hline
\end{tabular}
\end{minipage}
&
\begin{minipage}{.5\linewidth}
\begin{tabular}{cccccc}
\hline
\hline
Time & Vel. & Unc. & {H$\alpha$ EW} & {BIS} & {CCF FWHM} \\
{(JD$-$2440000)} & {(m/s)} & {(m/s)} & {(\AA)} & {(m/s)} & {(m/s)} \\
\hline
 18559.2685 &  5.829 & 0.340 & 1.9388 & -76.7 & 7645.4 \\
 18559.2728 &  5.485 & 0.336 & 1.9393 & -77.4 & 7645.8 \\
 18600.3138 & -6.818 & 0.323 & 1.9311 & -36.8 & 7613.4 \\
 18606.3136 &  1.378 & 0.281 & 1.9672 & -72.6 & 7574.6 \\
 18608.3141 &  1.033 & 0.293 & 1.9605 & -69.2 & 7580.9 \\
 18609.3433 &  1.598 & 0.335 & 1.9547 & -65.2 & 7586.8 \\
 18616.2515 & -4.564 & 0.314 & 1.9583 & -39.8 & 7605.9 \\
 18621.2651 &  1.294 & 0.291 & 1.9524 & -68.6 & 7563.7 \\
 18634.1729 &  3.776 & 0.275 & 1.9519 & -50.5 & 7588.9 \\
 18641.1586 &  4.078 & 0.356 & 1.9639 & -61.7 & 7627.1 \\
 18641.1641 &  2.774 & 0.329 & 1.9616 & -60.6 & 7622.7 \\
 18642.1850 &  1.429 & 0.307 & 1.9246 &  -    & -      \\
 18646.1772 &  3.214 & 0.367 & 1.9670 & -62.9 & 7629.2 \\
 18794.5362 & -8.143 & 0.372 & 1.9757 & -47.8 & 7574.5 \\
 18794.5393 & -7.524 & 0.409 & 1.9819 & -48.8 & 7575.8 \\
 18796.5347 & -6.448 & 0.375 & 1.9382 & -67.5 & 7558.2 \\
 18796.5379 & -7.336 & 0.383 & 1.9486 & -65.8 & 7554.9 \\
 18829.4906 &-10.890 & 0.354 & 1.9631 & -50.4 & 7580.1 \\
 18829.4933 & -9.922 & 0.378 & 1.9651 & -52.4 & 7581.5 \\
 18829.4970 &-10.811 & 0.380 & 1.9516 & -51.9 & 7579.1 \\
 18829.5012 &-10.594 & 0.384 & 1.9732 & -50.7 & 7580.8 \\
 18907.4231 &  2.478 & 0.343 & 1.9334 & -73.5 & 7596.3 \\
 18907.4275 &  2.647 & 0.316 & 1.9325 & -74.5 & 7595.1 \\
 19024.1584 & -8.145 & 0.314 & 1.9775 & -53.3 & 7592.6 \\
 19024.1668 & -8.378 & 0.328 & 1.9700 & -52.5 & 7590.1 \\
 19024.1751 & -8.263 & 0.336 & 1.9817 & -53.9 & 7594.0 \\
 19028.1551 &  0.813 & 0.351 & 1.9815 & -68.2 & 7572.7 \\
 19028.1580 & -0.437 & 0.346 & 1.9927 & -67.1 & 7573.0 \\
 19028.1613 & -0.152 & 0.354 & 1.9929 & -67.3 & 7572.6 \\
 19030.1878 & -3.715 & 0.436 & 2.0105 & -63.7 & 7570.4 \\
 19030.2002 & -3.288 & 0.356 & 2.0058 & -63.0 & 7563.4 \\
 19030.2118 & -3.930 & 0.368 & 2.0139 & -63.0 & 7566.5 \\
 19031.1579 & -2.632 & 0.338 & 1.9813 & -71.3 & 7558.4 \\
 19031.1624 & -1.930 & 0.337 & 1.9778 & -71.2 & 7554.3 \\
 19031.1666 & -2.136 & 0.350 & 1.9749 & -72.9 & 7549.4 \\
 19033.1566 &  2.298 & 0.394 & 2.0191 & -70.9 & 7573.2 \\
 19033.1585 &  2.511 & 0.389 & 2.0202 & -69.3 & 7574.0 \\
 19033.1622 &  1.977 & 0.361 & 2.0187 & -70.0 & 7572.5 \\
\hline
\end{tabular}
\end{minipage}
\end{tabular}
    \caption{RV measurements and activity indicators. Columns are: time of exposure, radial velocity following subtraction of the mean, uncertainty on velocity, Bisector Span, CCF Full-Width at Half-Maximum, and H$\alpha$ Equivalent width. Dashes (``-") indicate that the indicator could not be reliably measured.}
    \label{tab:rvsum}
\end{table*}

\begin{deluxetable}{cccccc}
\tablewidth{0pt}
\tablehead{
\colhead{Observing} & \colhead{} & \colhead{Date Range} & \colhead{Sigma} & 
\colhead{Seasonal Mean} & \colhead{Period} \\
\colhead{Season} & \colhead{$N_{obs}$} & \colhead{(HJD$-$2,400,000)} & 
\colhead{(mag)} & \colhead{(mag)} & \colhead{(days)} \\
\colhead{(1)} & \colhead{(2)} & \colhead{(3)} & \colhead{(4)} & 
\colhead{(5)} & \colhead{(6)}
}
\startdata
   1992--93   &  37 & 49095--49161 & 0.00435 & $-$0.65220(71) & 13.56 \\
   1993--94   &  96 & 49312--49519 & 0.00229 & $-$0.65817(23) & 16.70 \\
   1994--95   &  92 & 49687--49891 & 0.00464 & $-$0.65316(48) & 14.18 \\
   1995--96   & 102 & 50049--50256 & 0.00255 & $-$0.65933(25) & 13.82 \\
   1996--97   &  63 & 50404--50629 & 0.00284 & $-$0.65756(36) & 15.57 \\
   1997--98   &  90 & 50768--50991 & 0.00360 & $-$0.65699(37) & 16.55 \\
   1998--99   &  74 & 51123--51350 & 0.00397 & $-$0.65383(46) & 12.25 \\
   1999--00   &  82 & 51502--51710 & 0.00443 & $-$0.65858(49) & 16.46 \\
   2000--01   &  85 & 51866--52074 & 0.00711 & $-$0.64380(77) & 16.71 \\
   2001--02   &  72 & 52238--52461 & 0.00411 & $-$0.65103(48) & 15.73 \\
   2002--03   &  83 & 52595--52818 & 0.00309 & $-$0.65225(34) & 13.45 \\
   2003--04   &  84 & 52950--53190 & 0.00522 & $-$0.65473(57) & 16.39 \\
   2004--05   &  84 & 53311--53556 & 0.00443 & $-$0.65415(48) & 16.50 \\
   2005--06   &  92 & 53687--53905 & 0.00637 & $-$0.65090(66) & 16.75 \\
   2006--07   &  88 & 54046--54277 & 0.00509 & $-$0.65208(54) & 13.56 \\
   2007--08   &  93 & 54406--54636 & 0.00717 & $-$0.65223(74) & 16.70 \\
   2008--09   &  67 & 54831--54982 & 0.00711 & $-$0.64735(87) & 14.18 \\
   2009--10   &  67 & 55161--55368 & 0.00624 & $-$0.63311(76) & 13.82 \\
   2010--11   &  96 & 55529--55728 & 0.00735 & $-$0.64932(75) & 15.57 \\
   2011--12   & 125 & 55912--56097 & 0.00330 & $-$0.65570(30) & 16.55 \\
   2012--13   &  98 & 56256--56464 & 0.00513 & $-$0.65605(52) & 12.25 \\
   2013--14   & 108 & 56616--56825 & 0.00533 & $-$0.65378(51) & 16.46 \\
   2014--15   & 105 & 56988--57194 & 0.00289 & $-$0.65929(28) & 16.71 \\
   2015--16   & 146 & 57348--57558 & 0.00569 & $-$0.65075(71) & 15.73 \\
   2016--17   &  62 & 57707--57921 & 0.00288 & $-$0.65625(71) & 13.45 \\
   2017--18   &  77 & 58100--58281 & 0.00427 & $-$0.65808(49) & 16.39 \\
   2018--19   &  59 & 58597--58662 & 0.00309 & $-$0.65696(40) & 16.50 \\
   2019--20   &  63 & 58802--59022 & 0.00391 & $-$0.65783(49) & 16.75 \\
\enddata
    \caption{Summary of APT Photometric Observations for HD 101501}
    \label{tab:aptsum}
\end{deluxetable}

\end{document}